\documentclass[twocolumn,times,twocolappendix]{aastex631}
\newcommand*{\teff}{$T_{\rm eff}$\xspace}
\newcommand*{\logg}{$\log~g$\xspace}
\newcommand*{\feh}{[Fe/H]\xspace}

\newcommand*{\vt}{$\xi_{\rm t}$\xspace}
\newcommand*{\kms}{km s$^{-1}$\xspace}

\newcommand*{\msun}{$M_\odot$\xspace}

\newcommand*{\z}{$|Z|$\xspace}

\newcommand*{\rp}{$r$-process\xspace}

\newcommand*{\Ciso}{$^{12}\mathrm{C}/^{13}\mathrm{C}$\xspace}

\newcommand{\RomanNumeralCaps}[1] %% Roman number
    {\MakeUppercase{\romannumeral #1}} %% Roman number

\usepackage{xspace}
\usepackage{CJK}
\usepackage{amsmath}
\usepackage{url}
\usepackage{natbib}
\usepackage{tabularx}
\usepackage{color}
\usepackage{graphicx}
\usepackage{float}
\usepackage{epstopdf}
\usepackage{gensymb}
\usepackage{hyperref}
\usepackage{longtable}
\usepackage{subfigure}
\usepackage{verbatim}
\usepackage{orcidlink}

\hypersetup{
    citecolor=blue
}

\usepackage[T1]{fontenc}
\DeclareUnicodeCharacter{2212}{-}

\begin{document}

%\linenumbers

\shorttitle{A Detailed \rp Abundance Study of J1222}
\shortauthors{Jeong et al.}
\begin{CJK}{UTF8}{mj}

\title{The Actinide-Boost Star LAMOST J122216.85-063345.2: \\A Detailed $R$-process Abundance Study with Gemini-S/GHOST}
\author[0009-0009-7838-7771]{Miji Jeong}
\affiliation{Korea Astronomy and Space Science Institute, Daejeon 34055, Republic of Korea}
\affiliation{Department of Astronomy, Space Science, and Geology, Chungnam National University, Daejeon 34134, Republic of Korea}
\author[0000-0001-5297-4518]{Young Sun Lee (이영선)}
\affiliation{Department of Astronomy and Space Science, Chungnam National University, Daejeon 34134, South Korea; youngsun@cnu.ac.kr}
\author[0000-0003-4479-1265]{Vinicius M.\ Placco}
\affiliation{NSF NOIRLab, Tucson, AZ 85719, USA}
\author[0000-0002-5661-033X]{Yutaka Hirai}
\affiliation{Department of Community Service and Science, Tohoku University of Community Service and Science, 3-5-1 Iimoriyama, Sakata, Yamagata 998-8580, Japan}
\author[0000-0003-4573-6233]{Timothy C.\ Beers}
\affiliation{Joint Institute for Nuclear Astrophysics– Center for the Evolution of the Elements (JINA-CEE), USA}
\affiliation{Department of Physics and Astronomy, University of Notre Dame, Notre Dame, IN 46556, USA}
\author[0000-0003-2530-3000]{Jeong-Eun Heo}
\affiliation{Gemini Observatory/NSFʼs NOIRLab, Casilla 603, La Serena, Chile}

% Instrument Science Team
\author[0000-0002-9020-5004]{Kristin Chiboucas} 
\affiliation{Gemini Observatory/NSF’s NOIRLab, 670 N. A’ohoku Place, Hilo, HI, 96720, USA}
\author[0000-0001-9796-2158]{Emily Deibert} 
\affiliation{Gemini Observatory/NSFʼs NOIRLab, Casilla 603, La Serena, Chile}
\author[0000-0002-5227-9627]{Roberto Gamen} 
\affiliation{Instituto de Astrof\'isica de La Plata, CONICET--UNLP, and Facultad de Ciencias Astron\'omicas y Geof\'isicas, UNLP. Paseo del Bosque s/n, La Plata, Argentina}
\author[0000-0002-4641-2532]{Venu M. Kalari} 
\affiliation{Gemini Observatory/NSFʼs NOIRLab, Casilla 603, La Serena, Chile}
\author[0000-0002-5084-168X]{Eder Martioli} 
\affiliation{Laborat\'orio Nacional de Astrof\'isica, Rua Estados Unidos 154, 37504-364, Itajub\'a, MG, Brazil}
% Gemini Instrument Commissioning Team
\author{Ruben Diaz} 
\affiliation{Gemini Observatory/NSFʼs NOIRLab, Casilla 603, La Serena, Chile}
\author{Manuel Gomez-Jimenez}
\affiliation{Gemini Observatory/NSFʼs NOIRLab, Casilla 603, La Serena, Chile}
\author[0000-0001-7518-1393]{Roque Ruiz-Carmona} 
\affiliation{Gemini Observatory/NSFʼs NOIRLab, Casilla 603, La Serena, Chile}
\author{Greg Burley} 
\affiliation{NRC Herzberg Astronomy and Astrophysics Research Centre, 5071 West Saanich Road, Victoria, B.C., V9E 2E7, Canada}
\author[0000-0001-5528-7801]{J. Gordon Robertson}
\affiliation{Australian Astronomical Optics, Macquarie University, 105 Delhi Rd, North Ryde NSW 2113, Australia}
\affiliation{Sydney Institute for Astronomy, School of Physics, University of Sydney, NSW 2006, Australia}
\author[0000-0002-6633-7891]{Kathleen Labrie} 
\affiliation{Gemini Observatory/NSF’s NOIRLab, 670 N. A’ohoku Place, Hilo, HI, 96720, USA}
\author[0000-0003-1033-4402]{Joanna Thomas-Osip} 
\affiliation{Gemini Observatory/NSFʼs NOIRLab, Casilla 603, La Serena, Chile}
\begin{comment}
\end{comment}
%%%%%%%%%%%%%%%%%%%%%%%%%%%%%%%%%%%%%%%%%%%%%%%%%%%%%%%%%%%%%%%%%%%%%%%%%%%%%%%%%%%%%%%%%%%%%%%%%%%%%%%%%
%% Abstract %%
\begin{abstract}
We present a detailed chemical-abundance analysis of an actinide-boost ($\log\epsilon$\,(Th/Dy) = --0.74) star, LAMOST J122216.85-063345.2 (J1222), a very metal-poor ([Fe/H] = --2.45) halo star with moderate enhancement in rapid neutron-capture ($r$-)process elements ([Eu/Fe] = +0.61). From high-resolution spectra (R $\sim$ 55,000) taken with Gemini-S/GHOST, we determine the abundances for 47 elements, including thorium. The abundance pattern of J1222 is consistent with predicted nucleosynthetic yields from neutron star mergers (NSMs) and black hole-neutron star mergers (BH-NSMs), under specific ejecta conditions. Our kinematic analysis of J1222 indicates that it is a member of the I'itoi substructure. A comparative analysis of J1222 and seven other stars from the literature with similar dynamics to the I'itoi substructure exhibits a broad dispersion in $r$-process enrichment -- spanning non-enhancement ([Eu/Fe] $\leq$ +0.3), moderate enhancement (+0.3 $<$ [Eu/Fe] $\leq$ +0.7), strong enhancement ([Eu/Fe] $>$ +0.7), and actinide-boost stars (including one additional actinide-boost candidate newly recognized to be associated with I'itoi) -- suggesting a complex enrichment history shaped by multiple $r$-process events and inhomogeneous mixing. After exploring several astrophysical scenarios to explain the observed \rp abundances, we find that NSMs and BH-NSMs were likely the main contributors to the enrichment, while magneto-rotational supernovae (MR-SNe) may have played a secondary role in enriching some light \rp element-rich stars in the I'itoi substructure.
\end{abstract}

\keywords{Keywords: Unified Astronomy Thesaurus concepts: Chemical abundances (224);
Galaxy chemical evolution (580); Milky Way Galaxy (1054);
Stellar abundances (1577); Stellar populations (1622)}

%%%%%%%%%%%%%%%%%%%%%%%%%%%%%%%%%%%%%%%%%%%%%%%%%%%%%%%%%%%%%%%%%%%%%%%%%%%%%%%%%%%%%%%%%%%%%%%%%%%%%%%%%
%% INTRODUCTION
%% Section 1. Introduction %%
\section{Introduction} \label{sec:intro}
Roughly half of the elements heavier than $Z \sim 30$ are synthesized via the rapid neutron-capture process \citep[$r$-process;][]{burbidge_synthesis_1957, cameron_nuclear_1957}. Due to the extreme physical conditions required for \rp nucleosynthesis, several astrophysical sites have been proposed as its origin. These include neutron star mergers (NSMs) \citep[e.g.,][]{lattimer_black-hole-neutron-star_1974, lattimer_tidal_1976, goriely_r_2011, rosswog_long-term_2014, thielemann_neutron_2017}, black hole-neutron star mergers (BH-NSMs) \citep[e.g.,][]{meyer_decompression_1989, korobkin_astrophysical_2012, wanajo_actinide-boosting_2024}, magneto-rotational supernovae (MR-SNe) \citep[e.g.,][]{winteler_magnetorotationally_2012, nishimura_r_2015}, and collapsars \citep{surman_r_2008, siegel_collapsars_2019}. Among these, NSMs have gained compelling observational support, notably following the gravitational wave event (GW170817) and its associated kilonova \citep[e.g.,][]{abbott_gw170817_2017, cowperthwaite_electromagnetic_2017, drout_light_2017, kilpatrick_electromagnetic_2017, shappee_early_2017}. However, whether NSMs alone can fully account for various enhancements of \rp elements remains an open question.

Significant star-to-star variations in \rp abundances among metal-poor stars indicate rare, localized production events and inhomogeneous mixing in the early universe \citep[e.g.,][]{christlieb_hamburgeso_2004, honda_spectroscopic_2004, holmbeck_characterizing_2020}. A striking example is the ultra-faint dwarf galaxy Reticulum II, whose chemical composition appears to be dominated by a single \rp event \citep{ji_r-process_2016,ji_complete_2016,roederer_detailed_2016,ji_actinides_2018,ji_chemical_2023}.

Based on the level of enhancement in [Eu/Fe], metal-poor stars can be classified as either $r$-process-enhanced (RPE) ([Eu/Fe]\footnote{In this study, we adopt the standard abundance notation, where [A/B] is given by log(N$_{A}$/N$_{B}$)$_{*}$ $-$ log(N$_{A}$/N$_{B}$)$_{\odot}$, with N$_{A}$ and N$_{B}$ indicating the number densities of elements A and B. The absolute abundance of an element X is expressed as $\log\epsilon$(X) = $A$(X) = log(N$_{X}$/N$_{H}$) + 12, where N$_{X}$ and N$_{H}$ denote the number densities of element X and hydrogen, respectively.} $>$ +0.3) or non-RPE ([Eu/Fe] $\leq$ +0.3) stars. The RPE stars can be further sub-classified as moderately enhanced $r$-I (+0.3 $<$ [Eu/Fe] $\leq$ +0.7), or strongly enhanced $r$-II ([Eu/Fe] $>$ +0.7) stars (see \citealt{beers_discovery_2005, holmbeck_r-process_2020} for further details). Roughly one-third of the known RPE stars with measured thorium exhibit elevated Th/Eu abundance ratios, and these are referred to as ``actinide-boost'' stars \citep{hill_first_2002,honda_spectroscopic_2004,mashonkina_hamburgeso_2014,roederer_search_2014,siqueira_mello_high-resolution_2014,holmbeck_r-process_2018,placco_splus_2023}. Although NSMs, BH-NSMs, and MR-SNe have been proposed as potential sources of actinide production, the dominant formation pathway for actinides remains unclear.

Numerous RPE stars have been discovered in stellar substructures of the MW's halo, including Gaia-Sausage-Enceladus, Sequoia, and the Typhon stream \citep[e.g.,][]{matsuno_r-process_2021,matsuno_high-precision_2022-1, ji_chemical_2023}. These stars are believed to have originated in now-disrupted dwarf galaxies that were accreted into the halo during the hierarchical assembly of the Milky Way (MW). Their dynamical association with these substructures offers valuable insights into the spatial distribution of \rp production sites and the role of early mergers in shaping the chemical enrichment history of the MW.

We serendipitously discovered LAMOST J122216.85-063345.2 (hereafter, J1222) during our spectroscopic survey of very metal-poor (VMP; [Fe/H] $< -2.0$) stars selected from the Large Sky Area Multi-Object Fiber Spectroscopic Telescope (LAMOST) survey \citep{cui_large_2012}, as detailed in M. Jeong et al. (in preparation). A preliminary dynamical analysis suggested that J1222 is associated with the I'itoi substructure \citep{naidu_evidence_2020}. In this study, we conducted a detailed chemo-dynamical investigation of J1222, aiming to constrain the nucleosynthetic path of its neutron-capture elements and chemical-evolutionary history. By comparing its abundance pattern with RPE stars having similar dynamics to the I'itoi substructure selected from the Stellar Abundances for Galactic Archaeology (SAGA) database \citep{suda_stellar_2008, suda_stellar_2011}, we attempt to assess how its chemical and dynamical properties inform the chemical-evolutionary history of the I'itoi substructure, which can allow us to trace the chemical-evolution and early-enrichment history of the MW.

This paper is structured as follows. Section \ref{sec:2} describes the high-resolution spectroscopic observations and data reduction process. Section \ref{sec:3} outlines the determination of stellar atmospheric parameters and the measurement of elemental abundances. We present the chemical-abundance pattern of J1222, compare it with those of RPE and non-RPE stars in the SAGA database, and investigate its association with the I'itoi substructure in Section \ref{sec:4}. Section \ref{sec:5} discusses promising astrophysical scenarios that could account for the \rp enrichment of J1222, and explores the potential enrichment history of the I'itoi system. We summarize our findings in Section \ref{sec:6}.

%%%%%%%%%%%%%%%%%%%%%%%%%%%%%%%%%%%%%%%%%%%%%%%%%%%%%%%%%%%%%%%%%%%%%%%%%%%%%%%%%%%%%%%%%%%%%%%%%%%%%%%%%
%% BODY: Section 2 %%
\section{High-resolution Observations and Data Reduction}\label{sec:2}

\subsection{Target Observation with Gemini-S/GHOST}\label{sec:2.2}
To investigate the chemical evolution of known Galactic substructures, we previously conducted high-resolution spectroscopic observations of very metal-poor (VMP) candidates using Gemini Remote Access to CFHT ESPaDOnS Spectrograph \citep[GRACES;][]{navarro_graces_2014} on Gemini-N (M. Jeong, in preparation). Building on this work, we selected J1222 -- a bright, high-priority VMP target that was optimally visible during a System Verification (SV) run -- for follow-up observations with the Gemini High-resolution Optical Spectrograph \citep[GHOST;][]{ramsay_progress_2014,kalari_gemini_2024,mcconnachie_science_2024} on Gemini-S. GHOST is the newest optical high-resolution spectrograph on the Gemini-S telescope, and it has high sensitivity in the blue wavelength region where numerous absorption features due to neutron-capture elements can be observed.

We obtained high-resolution spectra of J1222 with the standard mode ($R \sim$ 55,000) of GHOST, operated in the 2-fiber mode with IFU1 targeting the star and IFU2 targeting the sky. We observed the star on May 10, 2023, as part of the GS-2023A-SV-101\footnote{\url{https://archive.gemini.edu/searchform/GS-2023A-SV-101-17/}} program. Both blue and red cameras were exposed for 3$\times$1500 s each. We used 1$\times$2 binning (spectral by spatial) for the observations. The observations were conducted under 70th-percentile image quality (IQ70), 70th-percentile cloud coverage (CC70), and 50th-percentile sky brightness (SB50).

\subsection{Data Reduction}\label{sec:2.3}
We reduced the 2D cross-dispersed echelle spectra using the GHOST-specific Data Reduction package, \texttt{GHOSTDR} v1.0.0 \citep{ireland_data_2018, hayes_ghost_2023}, implemented in the \texttt{DRAGONS} 3.0\footnote{\url{https://github.com/GeminiDRSoftware/DRAGONS}} software package \citep{labrie_dragonsquick_2023}. Standard calibration frames, including bias, flat-field, and arc lamp images, were used to reduce the raw data. The reduction steps included bias and flat-field corrections, wavelength calibration, sky subtraction, barycentric correction, and order combination.

The signal-to-noise ratio (S/N) of the extracted spectrum is 60 at 4000\,\AA, 200 at 5600\,\AA, and 260 at 6500\,\AA. The high S/N in the blue region highlights the exceptional data quality achieved with GHOST. This allows us to derive precise abundance ratios of neutron-capture elements, which are key to understanding the \rp enrichment history.

\begin{figure*}
    \centering
    \includegraphics[width=0.9\textwidth]{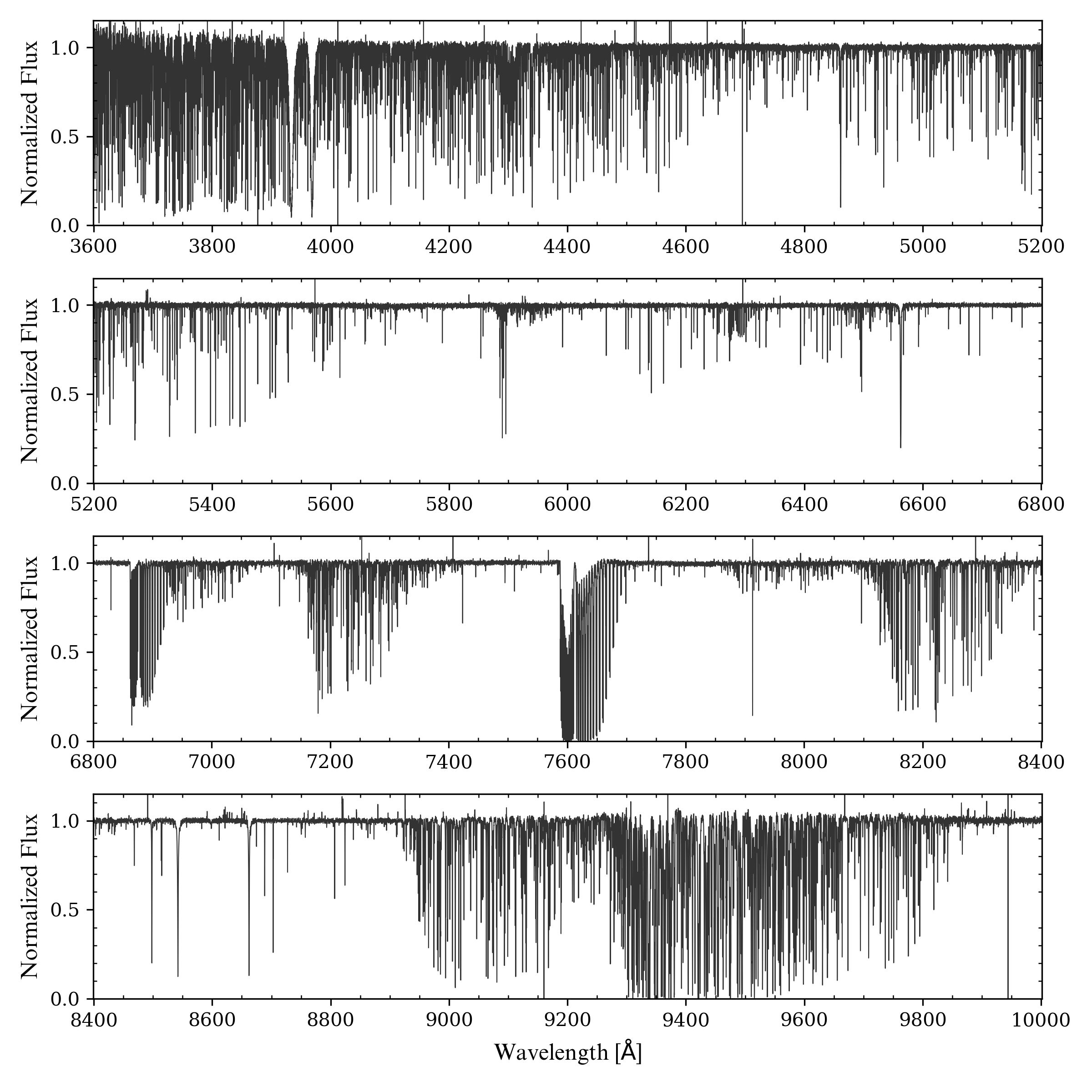}
    \caption{Normalized, order-combined spectrum of J1222, covering the wavelength range 3600--10000\,\AA.\\}
    \label{fig1}
\end{figure*}

After data reduction, we normalized each order individually. The spectra from multiple exposures were combined using a signal-weighted average to improve the S/N. The overlapping wavelength regions between adjacent orders were averaged by weighting the signal as well. Finally, the normalized and combined orders were merged to create a single, continuous spectrum for J1222. This spectrum was used for subsequent abundance analysis. Figure \ref{fig1} shows the final product of the spectrum of J1222, which covers the wavelength range from 3,600 to 10,000\,\AA. 

We determined radial velocity (RV) using a VMP template spectrum synthesized with \texttt{SYNTHE} \citep{alvarez_near-infrared_1998} from a KURUCZ model atmosphere. The RV measured from the red arm was 188.2$\pm$0.3 \kms, consistent with the low-resolution LAMOST (188.3 \kms) and moderate-resolution Gaia RVS (188.8 \kms). The consistency of RV measurements across different epochs suggests that J1222 is a single star. Heliocentric corrections were applied using the \texttt{astropy} package. The heliocentric RV ($V_{\rm helio}$) of J1222 is 169.5$\pm$0.6 \kms.

%% Table1 %%
\begin{deluxetable*}{lcccc}[htp]
\tablecaption{Astrometry, Stellar Parameters, and Orbital Parameters of LAMOST J122216.85-063345.2}
\label{tab:tab1}
\tablewidth{0pt}
\tabletypesize{\scriptsize}
\tablehead{
\colhead{Quantity} &
\colhead{Symbol} &
\colhead{Value} &
\colhead{Units} &
\colhead{Reference}
}
\startdata
\hline\hline
Right ascension           & $\alpha$ (J2000)    & ~~12:22:16.85         & hh:mm:ss.ss   & \citet{gaia_collaboration_gaia_2023}  \\
Declination               & $\delta$ (J2000)    & --06:33:45.2          & dd:mm:ss.s    & \citet{gaia_collaboration_gaia_2023} \\
Galactic longitude        & $\ell$              & 290.04                & degrees       & \citet{gaia_collaboration_gaia_2023} \\
Galactic latitude         & $b$                 & 55.57                 & degrees       & \citet{gaia_collaboration_gaia_2023} \\
Proper motion ($\alpha$)  & $\mu_{\alpha}$      & ~~2.512$\pm$0.020     & mas yr$^{-1}$ & \citet{gaia_collaboration_gaia_2023} \\
Proper motion ($\delta$)  & $\mu_{\delta}$      & --12.163$\pm$0.012    & mas yr$^{-1}$ & \citet{gaia_collaboration_gaia_2023} \\
Distance                  & $D$                 & 8.033$\pm$1.081       & kpc           &  Photometric  \\
Galactocentric coordinates (Cartesian) & ($X$, $Y$, $Z$) & (--6.63, --4.28, ~6.68 ) & kpc           & This study \\
Galactocentric distance   & $R$                 & 44.31$\pm$7.79        & kpc           & This study \\
\hline
Effective temperature     & \teff               & 4724$\pm$55       & K & This study \\
                          & \nodata             & 4596              & K & SSPP \citep{lee_segue_2008} \\
Log of surface gravity    & \logg               & $1.35^{+0.09}_{-0.12}$        & (cgs)         & This study \\
                          & \nodata             & 1.52                          & (cgs)         & SSPP \citep{lee_segue_2008} \\
Metallicity               & [Fe/H]              & --2.45$\pm$0.01     & \nodata       & This study \\
                          & \nodata             & --2.37              & \nodata       & SSPP \citep{lee_segue_2008} \\
Microturbulence velocity   & \vt                 & 2.0$\pm$0.2         & \kms\         & This study \\
\hline
$V$ magnitude             & $V$                 & 12.556$\pm$0.028      & mag           & APASS \citep{henden_vizier_2016} \\ %not V0
$K_{\rm s}$ magnitude         & $K_{\rm s}$             & 10.045$\pm$0.021      & mag           & 2MASS \citep{skrutskie_two_2006} \\ %not K0
Color excess              & $E(B-V)$            & 0.023                 & mag           & \citet{schlafly_measuring_2011} \\
\hline %not fraction of EBV
Radial velocity           & $V_{\rm rad}$  & 188.2$\pm$0.3        & \kms        & This study \\ 
Heliocentric radial velocity           & $v_{\rm helio}$             & 169.5$\pm$0.6        & \kms        & This study \\ %rv helio and err
Galactocentric velocity (Cartesian) & ($v_{\rm x}$, $v_{\rm y}$, $v_{\rm z}$) & (--343.2$\pm$36.6, --116.2$\pm$33.7, --99.7$\pm$30.3) & \kms\        & This study \\
Galactocentric velocity (Cylindrical) & ($v_{\rm R}$, $v_{\phi}$, $v_{\rm z}$) & (--225.2$\pm$9.4, --284.1$\pm$63.3, --99.7$\pm$30.3) & \kms\        & This study \\         
Integrals of motion       & ($J_{\rm R}$, $J_{\phi}$, $J_{z}$) & (2.1$\pm$3.7, 2.2$\pm$0.5, 0.8$\pm$0.2) $\times$ $10^{3}$& kpc~\kms\    & This study \\                         
Perigalactic distance        & $r_{\rm min}$      & 7.14$\pm$1.36 & kpc           & This study \\
Apogalactic distance         & $r_{\rm apo}$       & 44.48$\pm$7.79 & kpc           & This study \\
Maximum orbital distance from Galactic midplane &$z_{\rm max}$& 29.50$\pm$6.66 & kpc & This study \\
Total orbital energy      & $E_{\rm tot}$           & --0.963$\times$$10^5$ & km$^{2}$ s$^{-2}$ & This study \\
Orbital eccentricity      & $e$                 & 0.72$\pm$0.07 & \nodata       & This study \\
\hline\hline
\enddata   
\end{deluxetable*}

%two column
% FIGURE 1

%%%%%%%%%%%%%%%%%%%%%%%%%%%%%%%%%%%%%%%%%%%%%%%%%%%%%%%%%%%%%%%%%%%%%%%%%%%%%%%%%%%%%%%%%%%%%%%%%%%%%%%%%
%% BODY: Section 3 %%
\section{Determination of Stellar Atmospheric Parameters and Chemical Abundances} \label{sec:3}
From the GHOST spectrum of J1222, we determined stellar atmospheric parameters and elemental abundances by equivalent-width and spectral-synthesis analyses with \texttt{MOOG} \citep{sneden_carbon_1973}, under the assumption of one-dimensional local thermodynamic equilibrium (1D LTE). We adopted Solar photospheric abundances from \citet{asplund_chemical_2009}. The required Kurucz model atmospheres \citep{castelli_new_2003} were generated by interpolation with the \texttt{ATLAS9} code \citep{castelli_notes_1997} from a pre-computed model grid, which employs newly computed opacity distribution functions (ODFs) and updated abundances \citep{castelli_new_2003}.

\subsection{Estimation of Stellar Atmospheric Parameters}\label{sec:3.1}
The initial stellar parameters for J1222 were obtained by processing the LAMOST spectrum with the SEGUE Stellar Parameter Pipeline \citep[SSPP;][]{allende_prieto_segue_2008,lee_segue_2008,lee_segue_2008-1,lee_segue_2011,smolinski_segue_2011,lee_carbon-enhanced_2013,lee_sspp_2015}. The estimated initial parameters are \teff\ = 4596\,K, \logg\ = 1.52, and \feh\ = --2.37. By modifying and revising the methodology introduced in \citet{jeong_search_2023}, we developed an automated routine to estimate stellar parameters through iterative steps. At each iteration, new atmospheric models were generated based on the updated parameters, and the parameter derivation was repeated until the derived parameters did not change within a tolerance of $\Delta$\feh $<$ 0.02\,dex. This iterative process ensured self-consistent measurement of the stellar parameters. We briefly describe the determination of the stellar parameters below, and refer the interested reader to \citet{jeong_search_2023} for more details.

\subsubsection{Effective Temperature (\teff)}\label{sec:3.1.1}
The effective temperature (\teff) was determined iteratively, following the procedure described in \citet{jeong_search_2023}. We first used the initial stellar parameters (\teff, \logg, \feh) from the SSPP to select the appropriate $V-K$ color-\teff-[Fe/H] relations to use. Three estimates of \teff\ were obtained: one based on the SSPP \teff corrected through the empirical relation of \citet{frebel_deriving_2013}, and two from the $V-K$ relations for giants. We obtain \teff\ = 4724 $\pm$ 55\,K, with the uncertainty reflecting the standard deviation of the three estimates.

\subsubsection{Surface Gravity ($\log~g$)}\label{sec:3.1.2}
 Surface gravity (\logg) was determined iteratively using the isochrone-fitting method. The absolute $V$ magnitude ($M_{\rm V}$) was calculated using the de-reddened $V_{\rm 0}$ and the photometric distance. At each iteration, isochrones were generated using the updated \teff and [Fe/H], assuming a fixed $\alpha$-abundance of +0.4. For each of the four age assumptions (10, 11, 12, and 13 Gyr), we generated 100 random isochrones by drawing [Fe/H] values from a normal distribution \(\mathcal{N}([\rm Fe/\rm H],\ 0.2^2)\) to evaluate the uncertainties. Isochrones in the $M_{\rm V}$-\teff plane were used to determine whether J1222 is a dwarf or a giant, and the most likely classification was selected. The final \logg was determined to be $1.35^{+0.09}_{-0.12}$, with an uncertainty derived from the 16th and 84th percentiles of the \logg distribution.

\subsubsection{Metallicity ([Fe/H]) and Microturbulence Velocity (\vt)}\label{sec:3.1.3}
The metallicity ([Fe/H]) and microturbulence velocity (\vt) were also determined iteratively, with fixed \teff and \logg. The metallicity was first derived spectroscopically by measuring the equivalent widths (EWs) of 230 Fe~I lines in the wavelength range 3600--8800\,\AA. These derivations were performed using the updated model atmosphere at each iteration. The microturbulence velocity (\vt) was adjusted iteratively by minimizing the slope of the log (EWs) (of the Fe~I lines) against their reduced equivalent widths log (EWs/$\lambda$). The derived metallicity was [Fe/H] = --2.45$\pm$0.01. The uncertainty was calculated by considering the EW abundance uncertainty and the stellar parameter variation. (see Section \ref{sec:3.4}). The final microturbulence velocity (\vt) was 2.0$\pm$0.2 \kms. Table \ref{tab:tab1} lists the stellar parameters derived from the GHOST spectrum (first listed result) and by applying the SSPP (second listed result) to the low-resolution LAMOST spectrum \citep{lee_segue_2008}.

\vskip1.5cm
\subsection{Atomic Data For Equivalent width and Spectral Syntheses}\label{sec:3.3.1}
The chemical species used for EW analysis include \ion{Na}{1}, \ion{Mg}{1}, \ion{Si}{1}, \ion{Ca}{1}, \ion{Sc}{2}, \ion{Ti}{1}, \ion{Ti}{2}, \ion{Cr}{1}, \ion{Mn}{1}, \ion{Fe}{1}, \ion{Fe}{2}, and \ion{Ni}{1}. The EWs were measured by fitting Gaussian profiles with the \texttt{IRAF}\footnote{NOIRLab IRAF is distributed by the Community Science and Data Center at NSF NOIRLab, which is managed by the Association of Universities for Research in Astronomy (AURA) under a cooperative agreement with the U.S. National Science Foundation.} \texttt{splot} task \citep{tody_iraf_1986, tody_iraf_1993,fitzpatrick_iraf_2025}. These lines were selected from multiple sources (including \citealt{jang_graces_2023, jeong_search_2023, placco_metal-poor_2015, roederer_r-process_2018, roederer_r-process_2022}, and references therein). The log $gf$ values and excitation potentials for these lines were obtained from \texttt{LINEMAKE}\footnote{\url{https://github.com/vmplacco/linemake}} \citep{placco_linemake_2021,placco_linemake_2021-1}. To ensure reliability, we selected only well-isolated lines with minimal blending. This selection minimizes systematic errors in the EW measurements and leads to more accurate abundance determinations.

Spectral synthesis is crucial for determining the abundances of neutron-capture elements, as many spectral lines are weak or blended in metal-poor stars. We mitigated blending effects by modeling the full line profiles and achieved more reliable abundance measurements. We combined line data from \texttt{LINEMAKE} and \citet{roederer_r-process_2018}. Specifically, the species for neutron-capture elements considered in this study include \ion{Sr}{2}, \ion{Y}{2}, \ion{Zr}{2}, \ion{Mo}{1}, \ion{Ba}{2}, \ion{La}{2}, \ion{Ce}{2}, \ion{Pr}{2}, \ion{Nd}{2}, \ion{Sm}{2}, \ion{Eu}{2}, \ion{Gd}{2}, \ion{Tb}{2}, \ion{Dy}{2}, \ion{Ho}{2}, \ion{Er}{2}, \ion{Tm}{2}, \ion{Yb}{2}, \ion{Lu}{2}, \ion{Hf}{2}, \ion{Os}{1}, \ion{Ir}{1}, \ion{Pb}{1}, \ion{Th}{2}, and \ion{U}{2} which were adopted from \citet{roederer_r-process_2018}. The $\log~gf$ values and excitation potentials were taken from the \texttt{LINEMAKE} database. This combination provided accurate atomic data for reliable spectral synthesis, hence accurate abundance determination of neutron-capture elements. For lighter elements ($Z < 30$), including Li, C, N, [\ion{O}{1}], \ion{Al}{1}, \ion{V}{1}, \ion{V}{2}, \ion{Co}{1}, \ion{Cu}{1}, and \ion{Zn}{1}. we also adopted line data from \texttt{LINEMAKE}. Their abundances were derived through spectral synthesis, which allowed us to account for line blending and weak absorption features that could not be resolved through EW measurements.

In summary, we determined the abundances of 47 chemical elements. EW measurements were employed for unblended lines, while spectral synthesis was applied to neutron-capture elements (Z $\geq$ 30) and elements affected by significant blending or hyperfine structure effects. The isotopic ratios for all elements were adopted from \citet{sneden_neutron-capture_2008}, ensuring consistency with the Solar reference values. The full line list used in the abundance analysis, including wavelengths, excitation potentials, oscillator strengths, and derived abundances for each line, is presented in Table \ref{tab:linelist}.

% FIGURE 2
\begin{figure}[htp!]\vskip 0.2cm
    \includegraphics[width=1\linewidth]{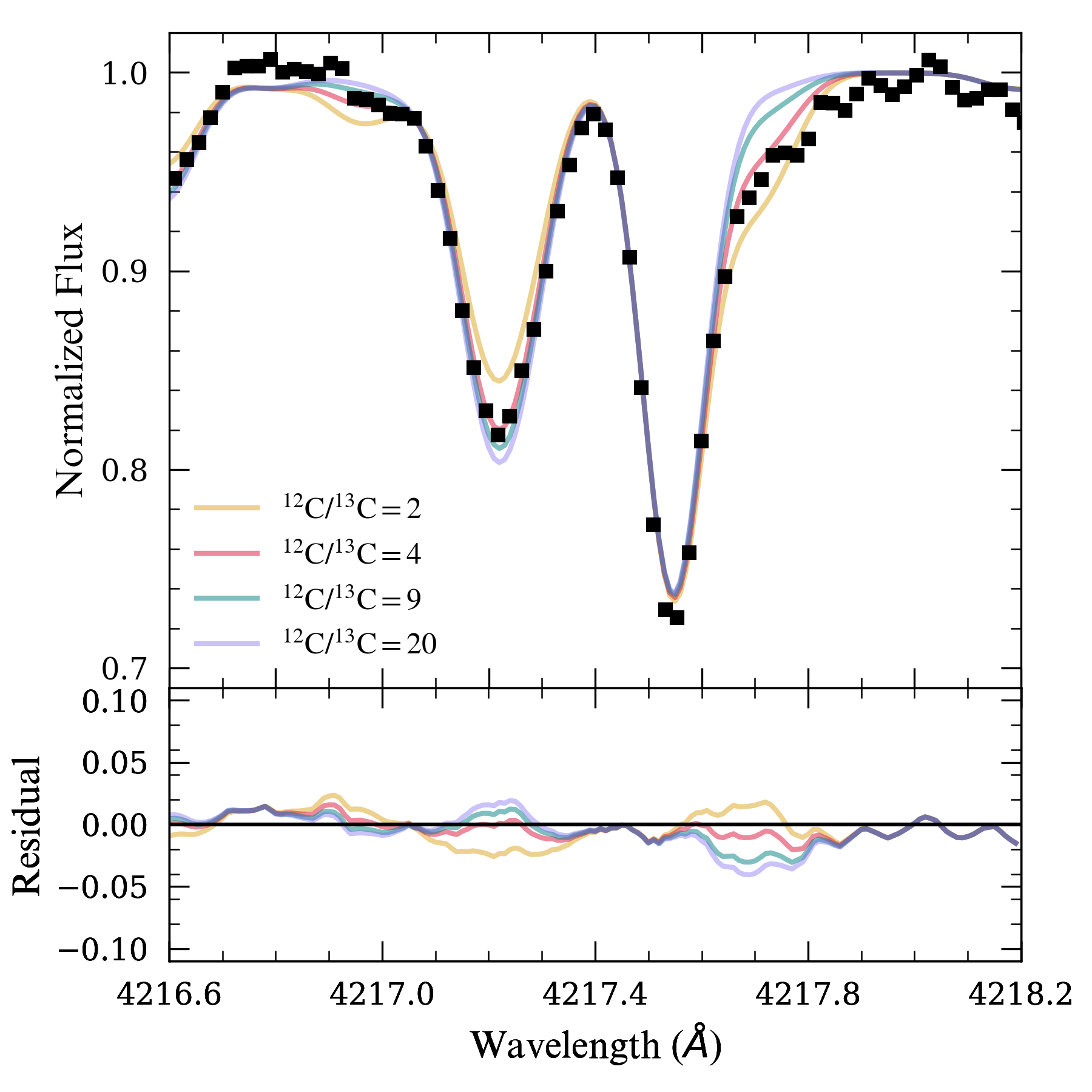}
    \caption{Top panel: Determination of carbon isotope ratio (\Ciso) for J1222. The black squares are the observed spectrum, and the red-solid line is the best-matched synthetic spectrum. The other colored solid lines show various spectral fits of \Ciso = 2, 4, 9, and 20, respectively. Bottom panel: Residual plot of the top panel.}
    \label{fig_c12c13}
\end{figure}

%\vskip 1cm
\subsection{Determination of Individual Elemental Abundances}\label{sec:3.3}
\subsubsection{Carbon and Nitrogen Abundances}\label{sec:3.3.2}
The carbon isotope ratio ($^{12}\mathrm{C}/^{13}\mathrm{C}$) for J1222 was estimated to be 4.0, based on the characteristics of CH around 4217\,\AA, as can be seen in Figure \ref{fig_c12c13}. In the figure, the observed spectrum is shown as filled squares, and the red-solid line represents the best-fitted model for ($^{12}\mathrm{C}/^{13}\mathrm{C}$) = 4.0. The lower panel of Figure \ref{fig_c12c13} shows the residuals between the observed spectrum and synthetic spectra for various isotope ratios. The low ratio indicates that a substantial portion of $^{12}\mathrm{C}$ was converted to $^{13}\mathrm{C}$ through internal mixing. 

\begin{figure*}[htp!]
    \centering
    \includegraphics[width=1\textwidth]{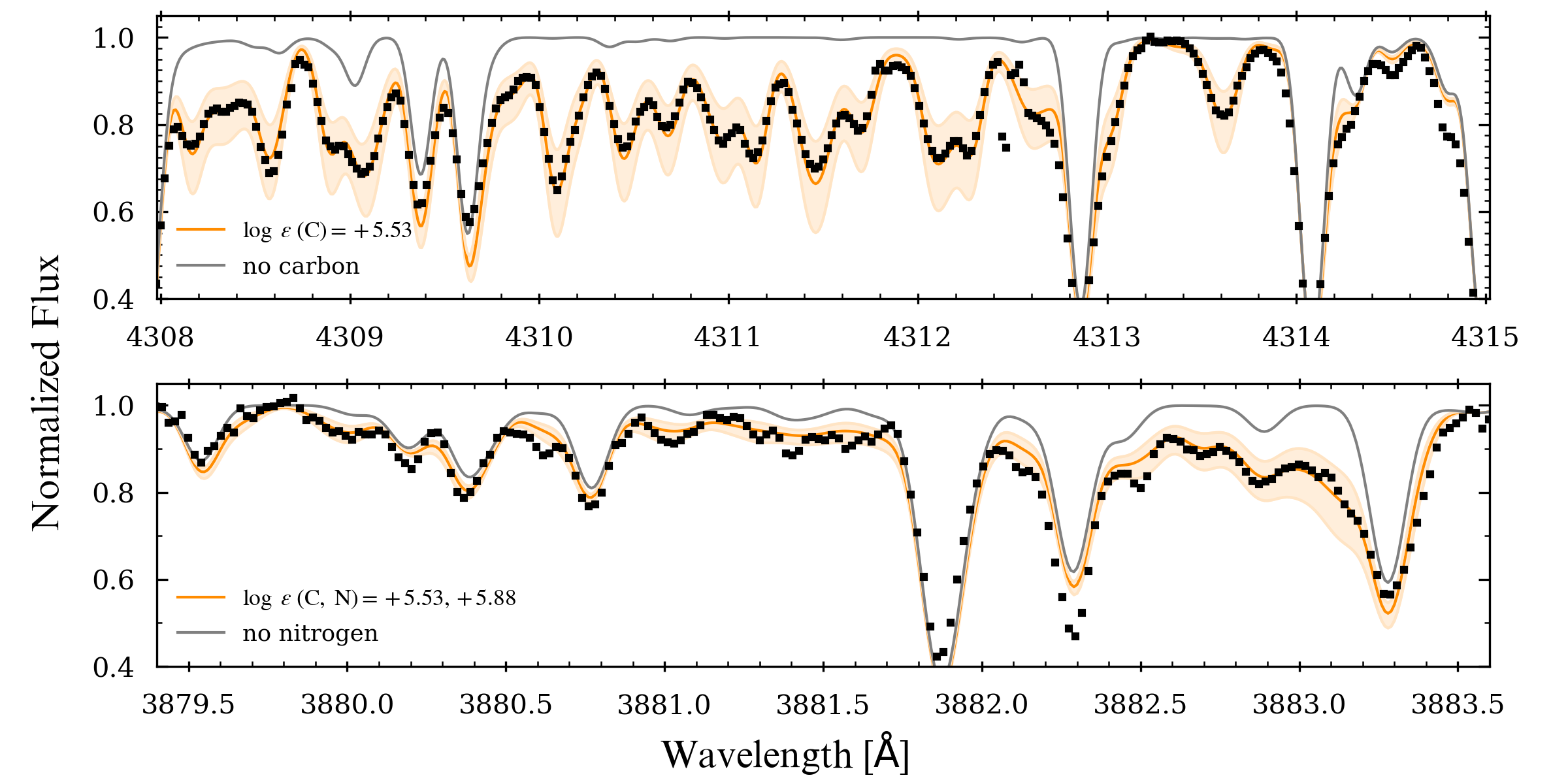}
    \caption{Observed spectrum (black dots) and synthetic spectra in the CH-$G$ band near 4308--4315\,\AA\ (top panel) and the CN band around 3879.5--3883.5\,\AA\ (bottom panel). The orange lines represent the best-fitted synthetic spectrum with carbon ($\log\epsilon$\,(C) = +5.53, top) and the combined carbon and nitrogen abundances ($\log\epsilon$\,(C, N) = +5.53, +5.88, bottom). In contrast, the gray lines show models without carbon or nitrogen. The orange-shaded region represents the uncertainty in the synthetic models, $\pm0.2$~dex.\\}
    \label{fig_ch_cn}
\end{figure*}

The carbon abundance was derived using spectral synthesis of the CH $G$-band around 4313\,\AA. The spectral synthesis fit to the CH $G$-band is shown in the upper panel of Figure \ref{fig_ch_cn}. A best-fit synthetic spectrum yielded $\log\epsilon$\,(C) = 5.53$\pm$0.27. We applied the correction described in \citet{placco_carbon-enhanced_2014} to take into account the effect of giant-branch evolution. The measured and corrected carbon abundances are listed in Table \ref{tab:tab2}.  

The nitrogen abundance was determined using the CN band near 3883\,\AA, fixing the carbon abundance at $\log\epsilon$\,(C) = 5.53, derived from the CH $G$-band. Spectral synthesis yielded $\log\epsilon$\,(N) = 5.88$\pm$0.26. The lower panel of Figure \ref{fig_ch_cn} shows the fit of the spectral synthesis to the CN band.

%two column
% FIGURE 3

\vskip 1cm
\subsubsection{Lithium}\label{sec:Li}
The lithium abundance was measured from the Li~I 6707.8\,\AA\ doublet. Since J1222 is an evolved red giant, lithium is expected to be strongly depleted as a consequence of the first dredge-up and subsequent extra-mixing processes along the red giant branch (RGB). Therefore, we adopt an upper limit of $\log\epsilon$\,(Li) $<$ $-0.60$. This is consistent with the low Li abundances typically observed before and after the red giant bump \citep{pilachowski_lidepletion_1993,gratton_rgb_mixing_2000}, where lithium is expected to be significantly reduced due to the first dredge-up and subsequent non-canonical extra-mixing \citep{charbonnel_Limixing_1995}.

\subsubsection{Light Elements (O to Ca)}\label{sec:3.3.3}
The abundances of light elements from O to Ca were primarily derived from EW measurements, except for Al, which required spectral synthesis as a result of line blending. The oxygen abundance was derived from the [O~I] forbidden line at 6300\,\AA, yielding  $\log\epsilon$\,(O) = 7.04$\pm$0.22. Sodium (Na~I) and magnesium (Mg~I) abundances were measured from two and twelve lines, respectively, resulting in $\log\epsilon$\,(Na) = 4.29$\pm$0.11 and $\log\epsilon$\,(Mg) = 5.62$\pm$0.05. The aluminum (Al~I) abundance was derived by spectral synthesis of a single line, yielding $\log\epsilon$\,(Al) = 4.10$\pm$0.23. Silicon (Si~I) and calcium (Ca~I) abundances were determined from six and twenty-seven lines, respectively, yielding $\log\epsilon$\,(Si) = 5.93$\pm$0.13 and $\log\epsilon$\,(Ca) = 4.23$\pm$0.06, respectively. The derived light-element abundances of J1222 indicate enhancements of the $\alpha$-elements, with [Mg/Fe] = +0.46, [Si/Fe] = +0.86, and [Ca/Fe] = +0.33.

\subsubsection{Iron-peak Elements (Sc to Zn)}\label{sec:3.3.4}
The abundances of iron-peak elements (Sc to Zn) were determined using a combination of EW measurements and spectral synthesis. The scandium abundance was derived from 9 Sc~II lines, yielding $\log\epsilon$\,(Sc) = 0.79$\pm$0.12. Titanium abundances were measured from 21 Ti~I and 31 Ti~II lines, resulting in $\log\epsilon$\,(Ti~I) = 2.66$\pm$0.05 and $\log\epsilon$\,(Ti~II) = 2.86$\pm$0.13, respectively. Vanadium (V~I and V~II) was measured using spectral synthesis of four lines in total, yielding $\log\epsilon$\,(V~I) = 1.36$\pm$0.18 and $\log\epsilon$\,(V~II) = 1.58$\pm$0.12. Chromium (Cr~I), manganese (Mn~I) and nickel (Ni~I) were all derived from the EW analysis, with mean abundances of $\log\epsilon$\,(Cr) = 2.97$\pm$0.04, $\log\epsilon$\,(Mn) = 2.77$\pm$0.09, and $\log\epsilon$\,(Ni) = 3.73$\pm$0.04. A total of 230 Fe~I and 22 Fe~II lines were analyzed, yielding $\log\epsilon$\,(Fe~I) = 5.06$\pm$0.01 and $\log\epsilon$\,(Fe~II) = 5.10$\pm$0.16. The excellent agreement between neutral and singly ionized iron supports the reliability of the derived [Fe/H]. The abundances of Co (Co~I), Cu (Cu~I), and Zn (Zn~I) were measured via spectral synthesis, resulting in $\log\epsilon$\,(Co) = 3.06$\pm$0.25, $\log\epsilon$\,(Cu) = 0.84$\pm$0.25, and $\log\epsilon$\,(Zn) = 2.12$\pm$0.15.

\subsubsection{Neutron-capture Elements (Sr to U)}\label{sec:3.3.5}
The neutron-capture elements (Sr to U) were measured using spectral synthesis. Multiple lines, including strong resonance and subordinate transitions, were used where available.

Light neutron-capture elements such as Sr, Y, and Zr were determined from \ion{Sr}{2}, \ion{Y}{2}, and \ion{Zr}{2} lines, with abundances of $\log\epsilon$\,(Sr) = 0.27$\pm$0.16, $\log\epsilon$\,(Y) = --0.42$\pm$0.16, and $\log\epsilon$\,(Zr) = 0.28$\pm$0.12, respectively. The molybdenum (Mo~I) abundance was derived from a single Mo~I line at 3864\,\AA, yielding an upper limit of [Mo/Fe] $<$ +0.30.

Among heavier elements, Ba and La were measured from 6 \ion{Ba}{2} and 9 \ion{La}{2} lines, yielding $\log\epsilon$\,(Ba) = --0.16$\pm$0.18 and $\log\epsilon$\,(La) = --1.13$\pm$0.12, respectively. Other rare-earth elements (Ce, Pr, Nd, Sm, Eu, Gd, Tb, Dy) showed reliable detections with moderate-to-high precision. Holmium (\ion{Ho}{2}), erbium (\ion{Er}{2}), thulium (\ion{Tm}{2}), and ytterbium (\ion{Yb}{2}) were also detected, with abundances of $\log\epsilon$\,(Ho) = --1.44$\pm$0.16 (2 lines), $\log\epsilon$\,(Er) = --0.85$\pm$0.20 (4 lines), $\log\epsilon$\,(Tm) = --1.55$\pm$0.10 (3 lines), and $\log\epsilon$\,(Yb) = --1.16$\pm$0.21 (1 line), respectively. Europium, a key tracer of \rp nucleosynthesis, was determined from seven Eu~II lines with a mean of $\log\epsilon$\,(Eu) = --1.32$\pm$0.21. 

Lutetium (Lu) and hafnium (Hf) were each measured from a single ionized line, yielding $\log\epsilon$\,(Lu) = --1.20$\pm$0.16 from the \ion{Lu}{2} 6221\,\AA\ line and $\log\epsilon$\,(Hf) = --1.10$\pm$0.16 from the \ion{Hf}{2} 3918\,\AA\ line. Osmium (\ion{Os}{1}) and iridium (\ion{Ir}{1}) were measured from single lines, with $\log\epsilon$\,(Os) = --0.45$\pm$0.28 and $\log\epsilon$\,(Ir) = --0.57$\pm$0.33, respectively. 

\begin{figure*}
    \centering
    \includegraphics[width=1\textwidth]{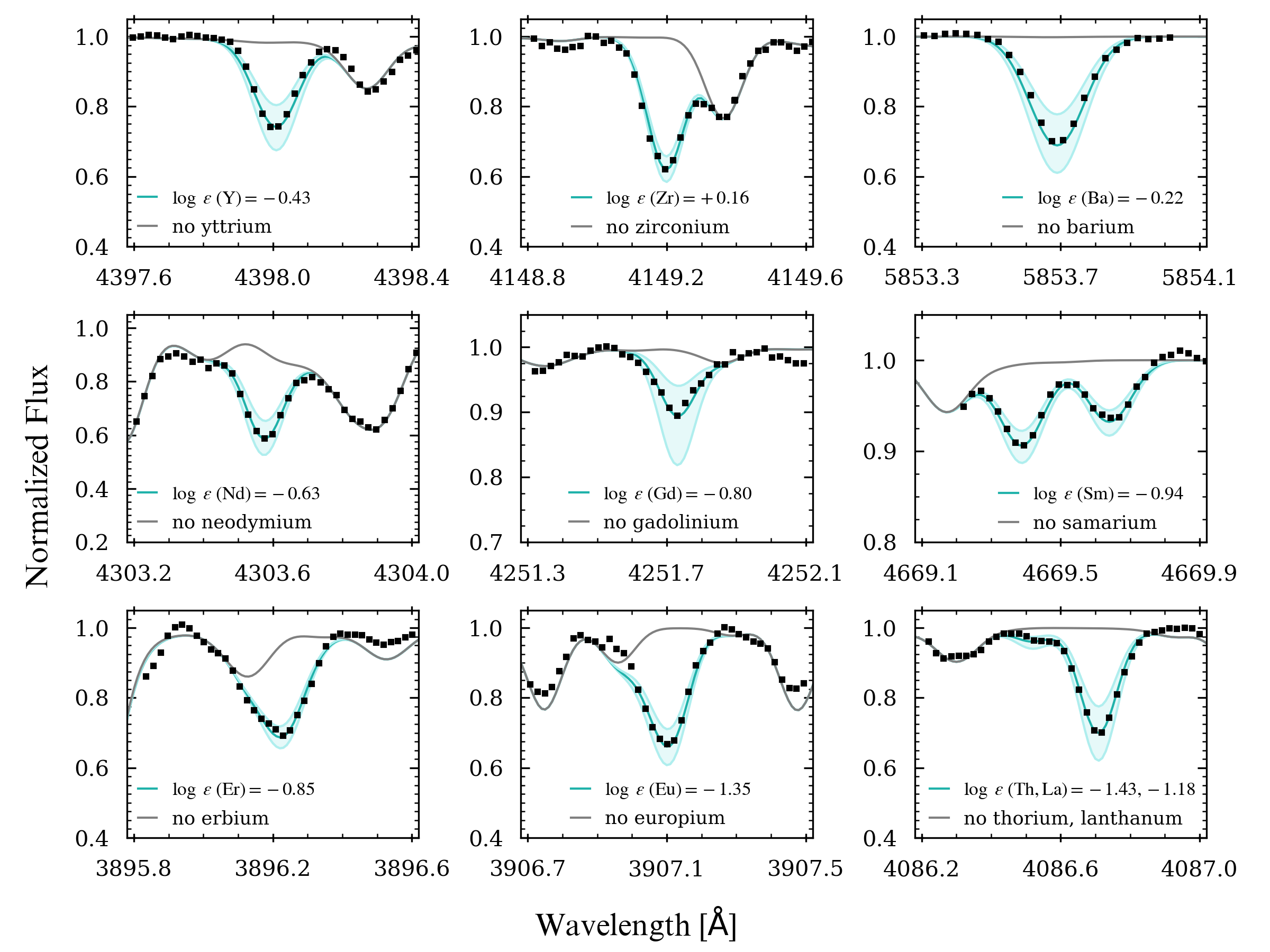}
    \caption{Spectral synthesis results derived for J1222 for important neutron-capture elements, from left to right and top to bottom, for Y, Zr, Ba, Nd, Gd, Sm, Er, Eu, La, and Th. Each panel shows the observed spectrum (black square), the best-fit model (cyan), and the no-abundance model (gray). The cyan-shaded regions represent the uncertainties ($\pm$0.2\,dex) in the synthetic fits.\\}
    \label{fig4}
\end{figure*}

Figure \ref{fig4} shows the spectral-synthesis results for key neutron-capture elements. Each panel compares the observed spectrum (black squares) with the best-fit synthetic model (green line) and a no-abundance model (gray line). The shaded regions represent uncertainties ($\pm$0.2\,dex) in the synthetic fit. 

We tried to detect the \ion{Pb}{1} line at 4058\,\AA, but no clear absorption feature was identified. Even when isotopic ratios and hyperfine splitting were included in the spectral synthesis, the line remained undetectable because of its intrinsic weakness and the S/N of the spectrum. A similar situation was reported by \citet{aoki_lead_2008}, who derived only an upper limit for HD~216143 using spectra of comparable resolution and stellar parameters, with S/N not much higher than ours. We therefore adopt a conservative upper limit of $\log\epsilon$\,(Pb) $<$ 0.0 for J1222. The actinide element thorium was measured from three Th~II lines, yielding $\log\epsilon$\,(Th) = --1.30$\pm$0.17.

For uranium (\ion{U}{2}), the 3859\,\AA\ line was not clearly detected, allowing only an upper limit to be derived. We considered all major blending species in this region and made slight adjustments to the abundances of Fe, V, and Sc to improve the local fit. However, the resulting line depth is only 1--2\% below the continuum, comparable to the residual scatter ($\sim$1--2\%) in the spectrum, making the feature indistinguishable from noise. Most reported uranium detections \citep{shah_uranium_2023} are based on spectra with S/N $\gtrsim$ 200 and R $\gtrsim$ 60,000, substantially higher S/N than ours. We therefore adopt a conservative upper limit of $\log\epsilon$\,(U) $<$ --1.99 for J1222.

\subsection{Abundance Uncertainties}\label{sec:3.4}
Depending on the measurement methods, we estimated the uncertainties of elemental abundances using either equivalent width (EW)-based analysis or a spectral synthesis (SYN)-based analysis. For EW-based elements, the total uncertainty was obtained by combining statistical and systematic uncertainties, whereas for SYN-based elements, conservative statistical uncertainties were estimated by examining how spectral-fitting residuals varied with changes in their abundance.

\subsubsection{Uncertainty of EW-based Abundances}\label{sec:3.4.1}
The EW-based abundance uncertainty ($\sigma_{\rm stat}$) was determined as the standard error of the mean abundance for a given element when more than two measurements were available. We calculated the line-to-line scatter ($\sigma$) of abundances from individual lines, then computed the standard errors for the number of measured elements by dividing the square root of the number of lines ($N$). In this way, we assume that the uncertainty includes local noise, continuum placement, and weak line blending effect, as reflected in line-to-line scatter. For elements with fewer than two measurements, the line-to-line scatter could not be reliably estimated. In such a case, we used the uncertainty of the \ion{Mg}{1} abundance, which has the largest scatter among EW-based analyses. However, we expect the actual uncertainty to be less than this.

The systematic uncertainties ($\sigma_{\rm sys}$) were estimated by varying the stellar parameters: $\pm$100\,K in $T_{\rm eff}$, $\pm$0.2\,dex in $\log g$, and $\pm$0.2 \kms\ in $\xi_{\rm t}$. The resulting abundance variations were combined in quadrature as their systematic uncertainty ($\sigma_{\rm sys}$) for each element. 

The total uncertainty ($\sigma_{\rm tot}$) for the EW-based abundances was obtained by adding $\sigma_{\rm stat}$ and $\sigma_{\rm sys}$ in quadrature, assuming that these two contributions are independent. The systematic uncertainties for EW-based elements are summarized in Table \ref{tab:tab_error}, while $\sigma_{\rm stat}$ and $\sigma_{\rm tot}$ are listed in Table \ref{tab:tab2}.

\vskip 1cm
\subsubsection{Uncertainty of SYN-based Abundances}\label{sec:3.4.2}
The SYN-based abundance uncertainty ($\sigma_{\rm stat}$) was estimated from the response of the spectral fitting residuals to variations in abundance. This reflects the fitting uncertainty arising from observational noise, including the effects of limited quality (S/N) and spectral resolution on continuum placement and line-profile fitting.

For every single line estimate, we generated synthetic spectra by changing the abundance by $\pm\Delta\log\epsilon$(X) ($\Delta$=0.1, 0.2, or 0.3\,dex) around the best-fit abundance, and RMS residuals between the observed and synthetic spectra were computed over the same wavelength range to mitigate the effects of nearby line mismatches and model limitations. For each line, the RMS residuals obtained from decreased and increased abundances relative to the best-fit residual were defined as $R^{-}$ and $R^{+}$, respectively. The effective residual was taken as $R_{\rm eff}$ = min($R^{-}$, $R^{+}$). The statistical uncertainty for each line was determined based on whether $R_{\rm eff}$ exceeded the predefined criteria. The criteria were empirically determined from the typical response of the fitting residuals to abundance variations of similar quality and resolution of the spectra.

To prevent uncertainties from becoming excessively large due to the sparse abundance sampling, an upper limit was imposed for cases with $\Delta$=0.3\,dex and for elements constrained only by upper limits. The resulting $\sigma_{\rm stat}$ includes uncertainties from fitting residuals, local continuum placement, line blending effects, and local noise. Therefore, systematic uncertainties were not separately estimated for the SYN-based abundances. When multiple lines were available for a given element, the median of the individual $\sigma_{\rm stat}$ was adopted as the final representative uncertainty. For elements measured from only one line, the corresponding $\sigma_{\rm stat}$ was adopted directly. For elements constrained only by upper limits, a conservative uncertainty of $\sigma_{\rm stat}$ = 0.2 \, dex was assigned when necessary. The statistical uncertainties are summarized in Table \ref{tab:tab2}. 

Moreover, we evaluated systematic uncertainties ($\sigma_{\rm sys}$) for SYN-based abundances by varying the adopted atmospheric parameters ($\pm$100\,K in $T_{\rm eff}$, $\pm$0.2\,dex in $\log g$, and $\pm$0.2 \kms\ in $\xi_{\rm t}$). While a fully line-by-line computation of $\sigma_{\rm sys}$ for all SYN-based abundances would be formally rigorous, the sensitivity of spectral lines to atmospheric parameters depends on their line-formation characteristics. We therefore grouped the SYN-based transitions by ionization stage and line strength, distinguishing molecular features, minority neutral species, majority ionized species, and heavy neutral lines. Similar sensitivity analyses have been applied in previous studies of metal-poor stars \citep[e.g.,][]{cayrel_first_2004,spite_sulfur_2011,heiter_gaiaeso_2021}.

Within each of these groups, we selected representative transitions from those used in the abundance analysis and recomputed their abundances under perturbed model atmospheres to quantify the impact of parameter variations. For each group, we adopted the largest abundance shift as a conservative estimate of $\sigma_{\rm sys}$ for other transitions with similar line-formation properties. For all other SYN-based species listed in Table \ref{tab:tab2}, $\sigma_{\rm sys}$ was assigned according to their corresponding line-formation group (molecular, minority neutral, majority ionized, or heavy neutral). Molecular features (CH, CN) showed the strongest temperature sensitivity, yielding $\sigma_{\rm sys}$ $\sim$ 0.21--0.22 dex. Minority neutral transitions (e.g., Al~I, Co~I, Zn~I) exhibited variations up to 0.15 dex, whereas majority ionized species (e.g., Eu~II, Zr~II) were comparatively stable, with $\sigma_{\rm sys}$ $\sim$ 0.06 dex. Heavy neutral lines (Os~I, Ir~I) showed larger responses, reaching up to 0.26 dex. All three lines of Th~II were calculated, yielding $\sigma_{\rm sys}$ $\sim$ 0.14 dex.

The adopted $\sigma_{\rm sys}$ values for each group are summarized in Table \ref{tab:tab_error}. In all cases, the largest abundance shift within the corresponding line-formation group was used, ensuring a conservative treatment of atmospheric parameter uncertainties. We propagated these systematic uncertainties into the total uncertainties ($\sigma_{\rm tot}$), which are listed in Table \ref{tab:tab2}.

%, we instead classified transitions according to their line-formation characteristics. Representative transitions were selected to different line-formation conditions, including molecular features, minority neutral species, majority ionized species, and heavy neutral transitions, and their abundance variations under modified model atmospheres were directly calculated \citep[e.g.,][]{cayrel_first_2004,spite_sulfur_2011,heiter_gaiaeso_2021}. 

%two column
% FIGURE 4

\begin{deluxetable}{lrrrcccc}%[htp]
\scriptsize
%\tabletypesize{\scriptsmall}
\tablewidth{0pc}
\tablecaption{Adopted Abundances and Statistical and Total Uncertainties for J1222}
\label{tab:tab2}
\tablehead{
\colhead{Species}                     & 
%\colhead{$\log\epsilon_{\odot}$\,(X)} & 
\colhead{$\log\epsilon$\,(X)}         & 
\colhead{$\mbox{[X/H]}$}              & 
\colhead{$\mbox{[X/Fe]}$}             & 
\colhead{$\sigma_{\rm stat}$}         & 
\colhead{$\sigma_{\rm tot}$}         & 
\colhead{$N$} &
\colhead{Method}
\\[-4pt]
\colhead{} &
\colhead{} &
\colhead{} &
\colhead{} &
\colhead{(dex)} &
\colhead{(dex)} &
\colhead{} &
\colhead{}
}       
\startdata
 \ion{Li}{1}     & $<$--0.60~~ & $<$--1.65 & $<$+0.80 & \nodata & \nodata & 1 & SYN\\
 \ion{C (CH)}{0} &     5.53~~ & $-$2.90 & $-$0.45 &    0.15 & 0.27 &   1 & SYN\\
 \ion{C}{0}$^{*}$&     5.68~~ & $-$2.75 & $-$0.30 &    0.15 & 0.27 &   1 & SYN\\ %CH G-band
 \ion{N (CN)}{0} &     5.88~~ & $-$1.95 &    +0.50 &    0.15 & 0.26 &   1 & SYN\\ %CN band
 $[$\ion{O}{1}$]$&     7.04~~ & $-$1.66 & +0.80 &    0.20 & 0.22 &   1 & SYN\\ %[OI] @6300A
 \ion{Na}{1}     &     4.29~~ & $-$1.96 &    +0.49 &    0.09 & 0.11 &   2 & EW\\ 
 \ion{Mg}{1}     &     5.62~~ & $-$1.99 &    +0.46 &    0.04 & 0.05 &  12 & EW\\
 \ion{Al}{1}     &     4.10~~ & $-$2.35 &    +0.10 &    0.20 & 0.23 &   1 & SYN\\ %Al II -> I
 \ion{Si}{1}     &     5.93~~ & $-$1.59 &    +0.86 &    0.11 & 0.13 &   6 & EW\\
 \ion{Ca}{1}     &     4.23~~ & $-$2.12 &    +0.33 &    0.02 & 0.06 &  27 & EW\\
 \ion{Sc}{2}     &     0.79~~ & $-$2.37 &    +0.08 &    0.03 & 0.12 &   9 & EW\\
 \ion{Ti}{1}     &     2.66~~ & $-$2.30 &    +0.15 &    0.02 & 0.05 &  21 & EW\\
 \ion{Ti}{2}     &     2.86~~ & $-$2.10 &    +0.35 &    0.02 & 0.13 &  31 & EW\\
 \ion{V}{1}      &     1.36~~ & $-$2.57 &   $-$0.12 & 0.10 & 0.18 &   2 & SYN\\ 
 \ion{V}{2}      &     1.58~~ & $-$2.35 &    +0.10 &    0.10 & 0.12 &   2 & SYN\\
 \ion{Cr}{1}     &     2.97~~ & $-$2.68 & $-$0.23 &    0.02 & 0.04 &   9 & EW\\
 \ion{Mn}{1}     &     2.77~~ & $-$2.67 & $-$0.22 &    0.09 & 0.09 &   7 & EW\\
 \ion{Fe}{1}     &     5.06~~ & $-$2.45 &    +0.00 & 0.01 & \nodata & 230 & EW\\
 \ion{Fe}{2}     &     5.10~~ & $-$2.41 &    +0.04 &    0.01 & 0.16 &  22 & EW\\
 \ion{Co}{1}     &     3.06~~ & $-$1.93 &    +0.52 &    0.20 & 0.25 &   3 & SYN\\
 \ion{Ni}{1}     &     3.73~~ & $-$2.50 & $-$0.05 &    0.02 & 0.04 &   24 & EW\\
 \ion{Cu}{1}     &     0.84~~ & $-$3.35 & $-$0.90 & 0.20 & 0.25 &   1 & SYN\\ %Oct22,2024
 \ion{Zn}{1}     &     2.12~~ & $-$2.43 &    +0.02 &    0.09 & 0.15 &   2 & SYN\\ 
 \ion{Sr}{2}     &     0.27~~ & $-$2.60 &   $-$0.15 &    0.15 & 0.16 &   2 & SYN\\
 \ion{Y}{2}      &  $-$0.42~~ & $-$2.63 & $-$0.18 &    0.15 & 0.16 &   6 & SYN\\
 \ion{Zr}{2}     &     0.28~~ & $-$2.30 &    +0.15 &    0.10 & 0.12 &   6 & SYN\\
 \ion{Mo}{1}     & $<$--0.27~~ & $<$--2.15 &    $<$+0.30 & \nodata & \nodata &   1 & SYN\\ %Mo II -> I
 \ion{Ba}{2}     &  $-$0.16~~ & $-$2.34 &    +0.09 &    0.17 & 0.18 &   6 & SYN\\
 \ion{La}{2}     &  $-$1.13~~ & $-$2.24 &    +0.21 &    0.10 & 0.12 &   9 & SYN\\
 \ion{Ce}{2}     &  $-$0.67~~ & $-$2.25 &    +0.20 &    0.10 & 0.12 &  8 & SYN\\
 \ion{Pr}{2}     &  $-$1.15~~ & $-$1.87 &    +0.58 &    0.10 & 0.12 &   8 & SYN\\
 \ion{Nd}{2}     &  $-$0.62~~ & $-$2.04 &    +0.41 &    0.20 & 0.21 &  23 & SYN\\
 \ion{Sm}{2}     &  $-$0.89~~ & $-$1.85 &    +0.60 &    0.10 & 0.12 &  21 & SYN\\
 \ion{Eu}{2}     &  $-$1.32~~ & $-$1.84 &    +0.61 &    0.20 & 0.21 &   7 & SYN\\
 \ion{Gd}{2}     &  $-$0.70~~ & $-$1.77 &    +0.68 &    0.20 & 0.21 &  14 & SYN\\
 \ion{Tb}{2}     &  $-$1.40~~ & $-$1.70 &    +0.75 &    0.20 & 0.21 &   1 & SYN\\
 \ion{Dy}{2}     &  $-$0.56~~ & $-$1.67 &    +0.78 &    0.10 & 0.12 &   8 & SYN\\
 \ion{Ho}{2}     &  $-$1.44~~ & $-$1.93 &    +0.52 &    0.15 & 0.16 &   2 & SYN\\
 \ion{Er}{2}     &  $-$0.85~~ & $-$1.77 &    +0.68 &    0.20 & 0.21 &   4 & SYN\\
 \ion{Tm}{2}     &  $-$1.55~~ & $-$1.65 &    +0.80 &    0.10 & 0.12 &   3 & SYN\\
 \ion{Yb}{2}     &  $-$1.16~~ & $-$2.00 &    +0.45 &    0.20 & 0.21 &   1 & SYN\\
 \ion{Lu}{2}     &  $-$1.20~~ & $-$1.30 &    +1.15 &    0.15 & 0.16 &   1 & SYN\\
 \ion{Hf}{2}     &  $-$1.10~~ & $-$1.95 &    +0.50 &    0.15 & 0.16 &   1 & SYN\\ 
 \ion{Os}{1}     &  $-$0.45~~ & $-$1.85 &    +0.60 &    0.10 & 0.28 &   1 & SYN\\ %Os II -> I
 \ion{Ir}{1}     &  $-$0.57~~ & $-$1.95 &    +0.50 &    0.20 & 0.33 &   1 & SYN\\ %Ir II -> I
 \ion{Pb}{1}     & $<$~0.00~~ & $<$--1.75 & $<$+0.70 & \nodata & \nodata &   1 & SYN\\
 \ion{Th}{2}     &  $-$1.30~~ & $-$1.32 &    +1.13 &    0.10 & 0.17 &   3 & SYN\\
 \ion{U}{2}      & $<$--1.99~~ & $<$--1.45 & $<$+1.00 &    \nodata & \nodata &   1 & SYN\\
\enddata
\tablecomments{*Abundance after application of the evolutionary corrections from \citet{placco_carbon-enhanced_2014}. Solar abundances ($\log\epsilon_{\odot}$\,(X)) are adopted from \citet{asplund_chemical_2009}. `SYN' indicates SYN-based analysis, whereas `EW' is the EW-based analysis.}
\end{deluxetable}

\begin{deluxetable}{
>{\centering\arraybackslash}p{1.1cm}
>{\centering\arraybackslash}p{1.1cm}
>{\centering\arraybackslash}p{1.1cm}
>{\centering\arraybackslash}p{1.1cm}
>{\centering\arraybackslash}p{1.1cm}
}
%\begin{deluxetable}{lcccc}%[htp]
%\begin{deluxetable}{p{1.1cm} p{1.1cm} p{1.1cm} p{1.1cm} p{1.1cm}}
\setlength{\tabcolsep}{6pt}
\tablewidth{0pc}
\tablecaption{Systematic Errors by Uncertainties of Atmospheric Parameters for Abundance Determinations of J1222}
\label{tab:tab_error}
\tablehead{
\colhead{Species} &
\colhead{$\Delta$\teff} &
\colhead{$\Delta$\logg} &
\colhead{$\Delta$\vt} &
\colhead{$\sigma_{\rm sys}$} 
%\colhead{$\sigma_{\rm tot}$}\\
\\[-4pt]
\colhead{} &
\colhead{$\pm$100\,K} &
\colhead{$\pm$0.2\,dex} &
\colhead{$\pm$0.2 \kms} &
%\colhead{} &
\colhead{(dex)}
}        
\startdata
\multicolumn{5}{c}{EW-based Species}\\
\hline
 \ion{Na}{1} &  0.05 &  0.02 & 0.05 &  0.07 \\
 ~\ion{Mg}{1} &  0.02 &  0.02 & 0.01 &  0.03 \\
 \ion{Si}{1}~~ &  0.05 &  0.02 & 0.03 &  0.06 \\
 \ion{Ca}{1} &  0.05 &  0.02 & 0.03 &  0.06 \\
 ~\ion{Sc}{2} &  0.08 &  0.08 & 0.02 &  0.12 \\
 \ion{Ti}{1}~~ &  0.02 &  0.01 & 0.04 &  0.05 \\
 \ion{Ti}{2} &  0.10 &  0.08 & 0.01 &  0.13 \\
 \ion{Cr}{1}~ &  0.01 &  0.02 & 0.03 &  0.03 \\
 ~\ion{Mn}{1} &  0.03 &  0.01 & 0.02 &  0.03 \\
 ~\ion{Fe}{2} &  0.13 &  0.10 & 0.02 &  0.16 \\
 \ion{Ni}{1} &  0.02 &  0.02 & 0.03 &  0.04 \\
 \hline
 \multicolumn{5}{c}{Representatives of SYN-based Species}\\
 \hline
 ~~~~\ion{C (CH)}{0} &  0.21 &  0.07 & 0.01 &  0.22 \\
 ~~~~\ion{N (CN)}{0} &  0.20 &  0.05 & 0.00 &  0.21 \\
 $[$\ion{O}{1}$]$~~~ &  0.06 &  0.05 & 0.01 &  0.08 \\
 \ion{Al}{1}~~~~ &  0.10 &  0.05 & 0.03 &  0.12 \\
 \ion{Co}{1}~~~ &  0.15 &  0.02 & 0.01 &  0.15 \\ 
 \ion{Zn}{1}~~~ &  0.04 &  0.02 & 0.00 &  0.04 \\
 \ion{Zr}{2}~~ & 0.05 &  0.03 & 0.00 &  0.06 \\
 \ion{Eu}{2}~~ &  0.05 &  0.04 & 0.00 &  0.06 \\
 \ion{Os}{1}~~~ & 0.24 &  0.08 & 0.00 &  0.26 \\
 \ion{Ir}{1}~~~~ & 0.15 &  0.01 & 0.00 &  0.15 \\
 \ion{Th}{2}~~ &  0.13 &  0.04 & 0.00 &  0.14 \\
\enddata
\tablecomments{Parameter sensitivities were evaluated for representative transitions used in the SYN-based abundance analysis, including CH-G band, the CN band, $[$\ion{O}{1}$]$ 6300 \AA, \ion{Al}{1} (3961 \AA), \ion{Co}{1} (4045 \AA), \ion{Zn}{1} (4722 \AA), \ion{Zr}{2} (4208 \AA), \ion{Eu}{2} (4129 \AA), \ion{Os}{1} (4420 \AA), \ion{Ir}{1} (3800 \AA), and \ion{Th}{2} (4019, 4086, 4094 \AA). $\sigma_{\rm sys}$ for other SYN-based elements were assigned according to the corresponding line-formation group described in Section \ref{sec:3.4.2}}
\end{deluxetable}

%%%%%%%%%%%%%%%%%%%%%%%%%%%%%%%%%%%%%%%%%%%%%%%%%%%%%%%%%%%%%%%%%%%%%%%%%%%%%%%%%%%%%%%%%%%%%%%%%%%%%%%%%
%% BODY: Section 4 %%
\section{Results} \label{sec:4}
Although we have measured abundances for a wide range of elements, including light and iron-peak elements, we do not report their detailed abundance patterns here. Instead, we focus on the behavior of the neutron-capture elements, as our primary interest is the \rp\ enrichment history of J1222. Our derived abundances for the first-, second-, and third-peak neutron-capture elements confirm that J1222 is moderately enriched across a broad range of neutron-capture elements. In this section, we present the neutron-capture abundance pattern and discuss it further in Section~\ref{sec:5.1}.

\subsection{$R$-process Enhancement and the Actinide-boost Nature of J1222}\label{sec:4.1}
\begin{figure*}
    \centering
   \includegraphics[scale=0.8]{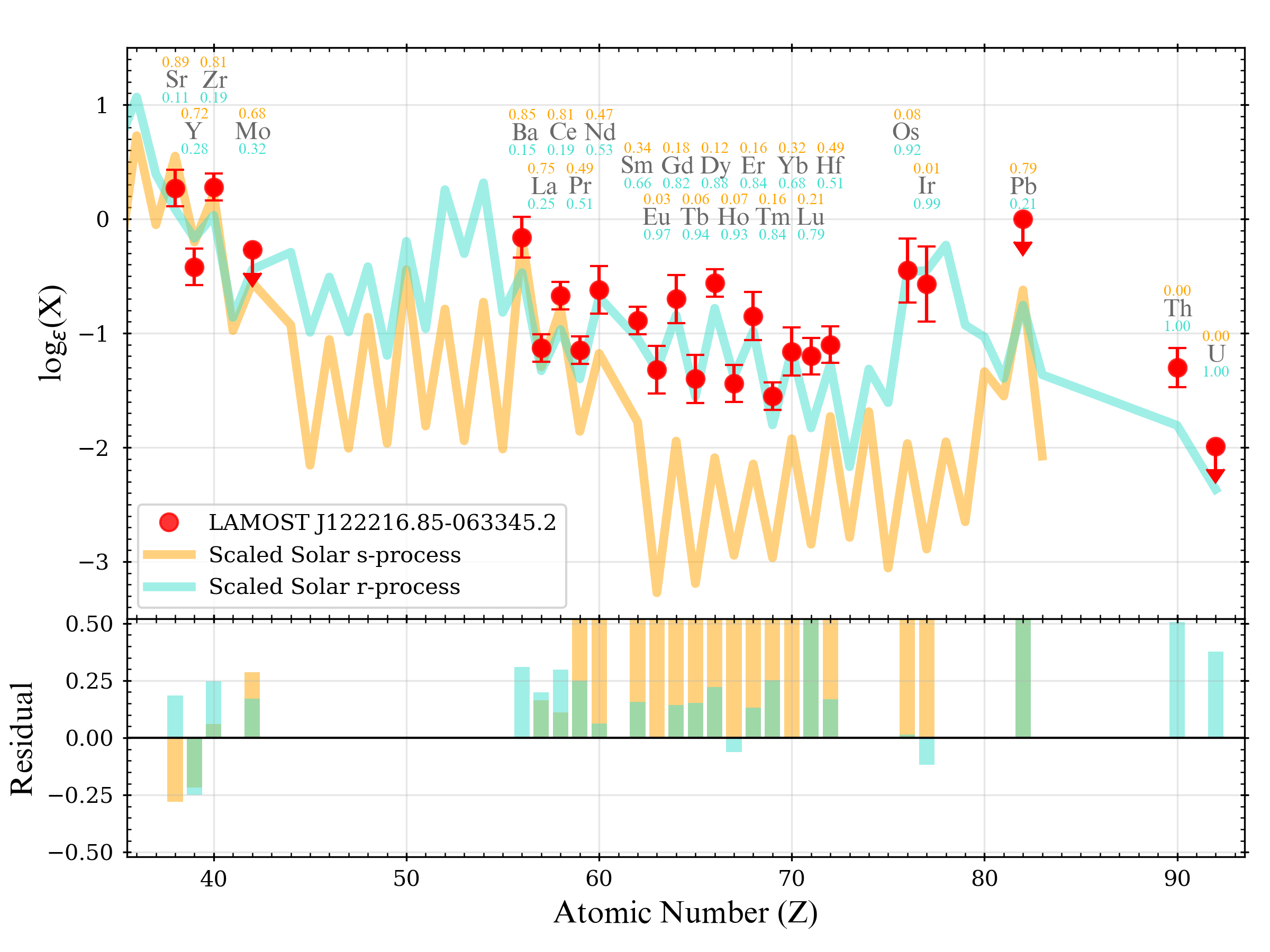}
    \caption{Top panel: Abundance ratios ([X/Fe]) of J1222 for heavier elements (Z $\geq$ 35), as a function of atomic number (Z). The red dots represent the measured abundances with their uncertainties (see Table \ref{tab:tab2}). The orange line shows the scaled-solar $s$-process pattern, normalized to the Ba abundance of J1222, while the blue line represents the scaled-solar \rp pattern, normalized to the Eu abundance of J1222. The Solar reference values are taken from \citet{asplund_chemical_2009}, and the $s$- and \rp contributions for each element are based on Table 5 from \citet{burris_neutron-capture_2000}. The numbers around each element symbol are the production fraction of $s$- (orange) and \rp (green) from \citet{burris_neutron-capture_2000}. Bottom panel: Residuals between the observed abundances of J1222 and the 
    scaled-solar $s$- and \rp patterns ($[\rm X/\rm Fe] - [\rm X/\rm Fe]_{s}$ or $[\rm X/\rm Fe] - [\rm X/\rm Fe]_{r}$).\\}
    \label{fig_solar}
\end{figure*}

The heavy-element ($Z > 30$) abundance pattern of J1222 was examined by comparing it to the scaled-Solar \rp and $s$-process patterns, to identify the primary enrichment source(s). To accomplish this, the abundances of Ba and Eu were used to produce the scaled-Solar $s$- and \rp patterns, following the fractional contributions proposed by \citet{burris_neutron-capture_2000}. Figure \ref{fig_solar} presents this comparison, with the upper panel showing the observed elemental abundances (red dots) alongside the scaled Solar $s$- and \rp patterns (orange and cyan, respectively). The lower panel shows the deviations from the $s$- and \rp trends. While most neutron-capture elements follow the scaled-Solar \rp pattern, some elements, such as Sr, Y, and Mo, exhibit offsets, suggesting additional nucleosynthetic pathway(s) beyond a pure \rp. 

Although the heavy-element pattern generally matches the scaled-solar \rp\ trend, we applied the classification scheme from \citet{frebel_nuclei_2018} to assess possible contributions from other nucleosynthesis sources, such as the weak \rp. In this scheme, limited-$r$ ($r_{\rm lim}$) stars are defined with [Eu/Fe] $\leq$ +0.3, [Sr/Ba] $>$ +0.5, and [Sr/Eu] $>$ 0.0. These criteria are intended to identify stars that exhibit light neutron-capture enhancement without significant enrichment in heavy \rp\ elements such as Eu. However, J1222 has [Sr/Ba] = --0.24 and [Sr/Eu] = --0.76, thus it does not meet the classification criteria for a limited-$r$ star. Instead, with [Eu/Fe] $>$ +0.3, it is classified as a $r$-I star.

Another possible source of enrichment is contamination from asymptotic giant branch (AGB) stars. To evaluate the potential $s$-process contribution from AGB stars, we examined [Ba/Eu] and [La/Eu], because if AGB stars played a significant role, these values would be positive. However, J1222 exhibits [Ba/Eu] = --0.52 and [La/Eu] = --0.40, indicating a minimal contribution from the $s$-process. In addition, AGB nucleosynthesis typically yields low [Sr/Ba] ratios as a result of enhanced Ba production via the $s$-process. The even lower [Sr/Ba] observed in J1222 ([Sr/Ba] = --0.24) falls below the typical AGB $s$-process range. We directly compared the observation with AGB model predictions using the \texttt{FRUITY}\footnote{\url{https://fruity.oa-teramo.inaf.it/modelli.pl}} database \citep{cristallo_evolution_2011}. At $Z$ = 0.001, AGB models for 1.3, 2.0, and 5.0 \msun\ stars predict [Ba/Eu] values of +0.67, +0.81, and +0.32, respectively. At $Z$ = 0.0001 with [$\alpha$/Fe] = 0.05, the predicted [Ba/Eu] ranges from +0.78, +0.77, and +1.19. In contrast, the observed [Ba/Eu] = --0.52 is significantly lower than all the predictions of the models, confirming that AGB stars did not play a significant role in the chemical enrichment of J1222.

To determine whether or not J1222 is an actinide-boost star, we adopted the classification criterion from \citet{holmbeck_actinide_2019}, which defines actinide-boost stars as those with $\log\epsilon$\,(Th/Dy) $>$ --0.9. J1222 meets this criterion, with $\log\epsilon$\,(Th/Dy) = --0.74$\pm$0.21. Furthermore, Figure \ref{fig_solar} shows that its Th abundance exceeds the scaled-solar \rp pattern by more than 0.5\,dex. This confirms J1222 as a new actinide-boost star, suggesting that it was not formed in a classical \rp site, but rather in an environment involving exceptionally high neutron fluxes. The $r$-I classification, low $s$-process contribution, and actinide-enhancement signatures imply that J1222 experienced a distinctive \rp pathway, shaped by extreme neutron flux environments. We further explore this possibility in Section \ref{sec:5}.

\begin{figure}[htp!]
    \includegraphics[width=1\linewidth]{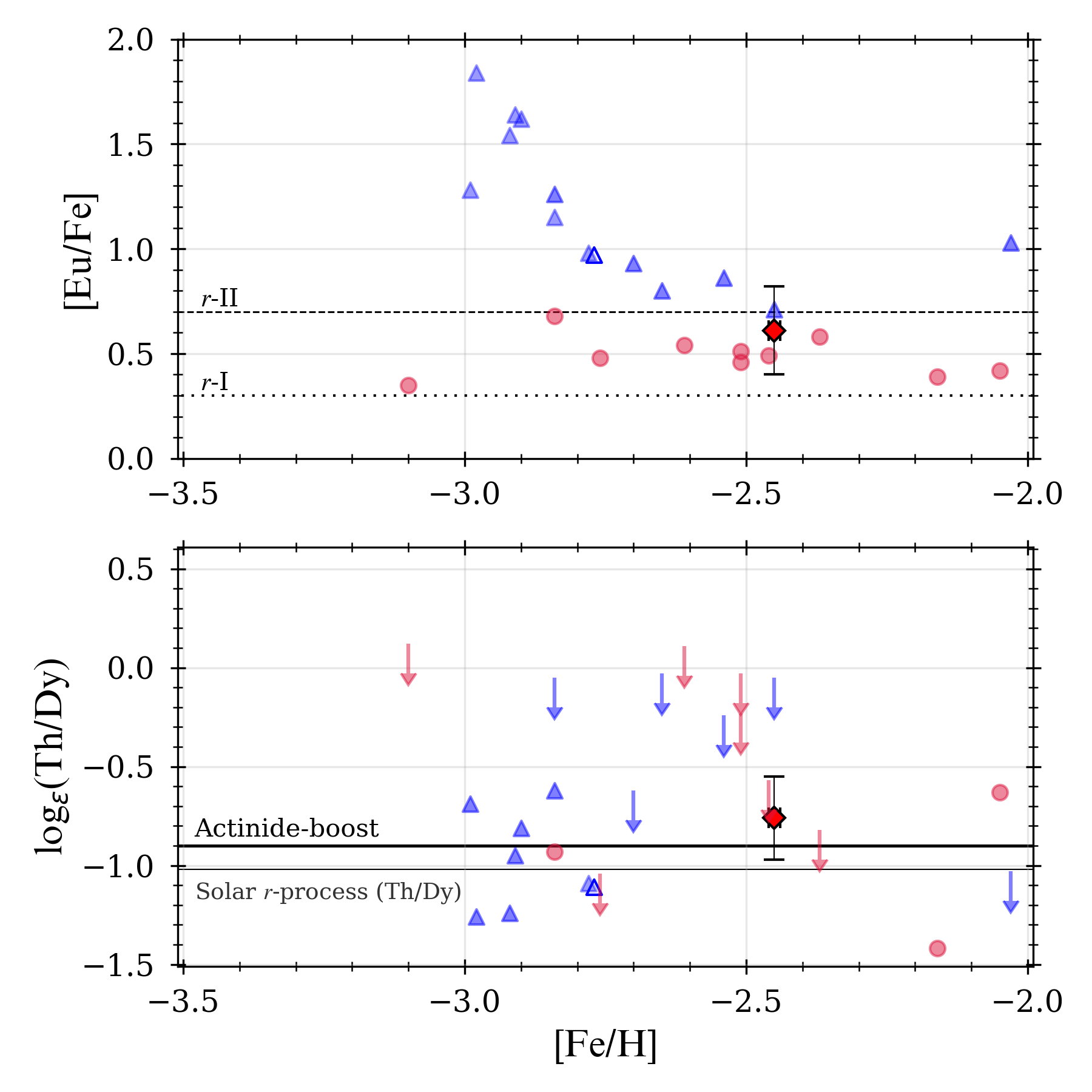}
    \caption{Top panel: [Eu/Fe] as a function of [Fe/H] for \rp-enhanced stars with available Th and Dy measurements. Red circles and blue triangles denote $r$-I and $r$-II stars, respectively. The horizontal lines indicate the $r$-I (+0.3 $\leq$ [Eu/Fe] $<$ +0.7; dotted) and $r$-II ([Eu/Fe] $\geq$ +0.7; dashed) boundaries. J1222 is highlighted with a diamond symbol and its measurement uncertainty. Bottom panel: $\log\epsilon$\,(Th/Dy) as a function of [Fe/H] for the same comparison sample. Filed symbols represent stars with detections of both Th and Dy, while arrows show upper limits in $\log\epsilon$\,(Th/Dy). The horizontal line at $\log\epsilon$\,(Th/Dy) $>$ --0.9, following \citet{holmbeck_actinide_2019}, indicates the actinide-boost criterion. The lower horizontal line marks the solar \rp Th/Dy ratio, $\log\epsilon$\,(Th/Dy) $\approx$ --1.02. The comparison data are taken from the literature \citep {honda_spectroscopic_2004,frebel_he15230901_2007,lai_mpstar_2008,mashonkina_hamburgeso_2010,mashonkina_hamburgeso_2014,roederer_search_2014,siqueira_mello_high-resolution_2014,placco_rave_2017,holmbeck_r-process_2018,sakari_rpa_2018}.}
    \label{fig_r_position}
\end{figure}

We compare J1222 with \rp enhanced stars for which both Th and Dy measurements are available to place its actinide-boost signature in the context of previously studied $r$-process enhanced stars. In the top panel of Figure \ref{fig_r_position}, J1222 lies in the canonical $r$-I regime and is indistinguishable from other $r$-I stars based on Eu abundance alone. In the bottom panel of Figure \ref{fig_r_position}, the comparison sample consists of 11 stars with detections of both Th and Dy and 13 stars with upper limits on Th (while Dy is detected). The lower horizontal line marks the solar \rp Th/Dy ratio ($\log\epsilon$\,(Th/Dy) $\approx$ --1.02), derived from the adopted solar abundances \citep{asplund_chemical_2009} and \rp fractions \citep{burris_neutron-capture_2000}. Within this sample, actinide-boost stars are found in both $r$-I and $r$-II classes. This indicates that actinide enhancement does not correspond solely to Eu enhancement. 

More generally, actinide enhancement appears to reflect variation in the physical conditions of the \rp, such as differences in the neutron-richness of ejecta \citep[e.g.,][]{holmbeck_actinide_2019,eichler_acdinide_2019}. Similar behavior has been reported in different halo environments \citep[e.g.,][]{placco_splus_2023,lin_actinide_gse_2025}. This comparison is limited to stars with measurable Th and Dy abundances, and is therefore subject to selection effects. J1222 provides an example of how Eu abundance alone does not fully characterize actinide behavior. Detailed comparisons of individual elemental abundances are presented in Section \ref{sec:4.2}.

%two column
% FIGURE 5

% FIGURE X (style2)

\vskip 1cm
\subsection{Constraints on U/Th Nuclear Chronometric Ages}\label{sec:4.1a}
Age estimates derived from standard cosmo-chronometers, such as Th/X or U/X (where X represents stable \rp elements such as Eu), often yield negative or nonphysical results in actinide-boost stars, because actinide overproduction hinders the assumption of a universal initial production ratio (PR). In contrast, the U/Th ratio is a more reliable chronometer for them because both uranium and thorium are actinides synthesized through similar nucleosynthetic pathways. Consequently, the U/Th ratio is less sensitive to variations in the overall \rp enrichment history than chronometers that involve stable \rp elements. 

The application of the U/Th chronometer implicitly assumes that the observed actinide abundances are dominated by a single actinide-producing \rp source, even if additional enrichment episodes could contribute to lighter \rp elements. Using the \ion{U}{2} 3859\,\AA\ line, we placed an upper limit on the uranium abundance and derived $\log\epsilon$\,(U/Th) $<$ --0.69 for J1222. Because the uranium abundance is constrained only by an upper limit, the resulting U/Th chronometric age cannot be considered a precise measurement with a well-defined uncertainty. Instead, it provides a conservative estimate of the time that has passed since the \rp enrichment event.

To estimate a U/Th chronometric constraint, we applied Equation (3) from \citet{cayrel_uranium_2001} and adopted several initial production ratios of $\log\epsilon$\,(U/Th)$_{0}$ from the literature, as summarized by \citet{roederer_r-process_2024}: --0.26 \citep{cowan_pr_1999}, --0.22 \citep{cowan_pr_2002}, --0.20 \citep{schatz_pr_2002}, and --0.28 \citep{kratz_pr_2007}.  Consequently, U/Th chronometer provides lower limits on the age, ranging from $\ge$8.89 to $\ge$10.63 Gyr depending on the adopted production ratio.

\subsection{Comparison with Milky Way Non-RPE and RPE Stars}\label{sec:4.2}
\begin{figure*}
    \centering
    \includegraphics[scale=0.8]{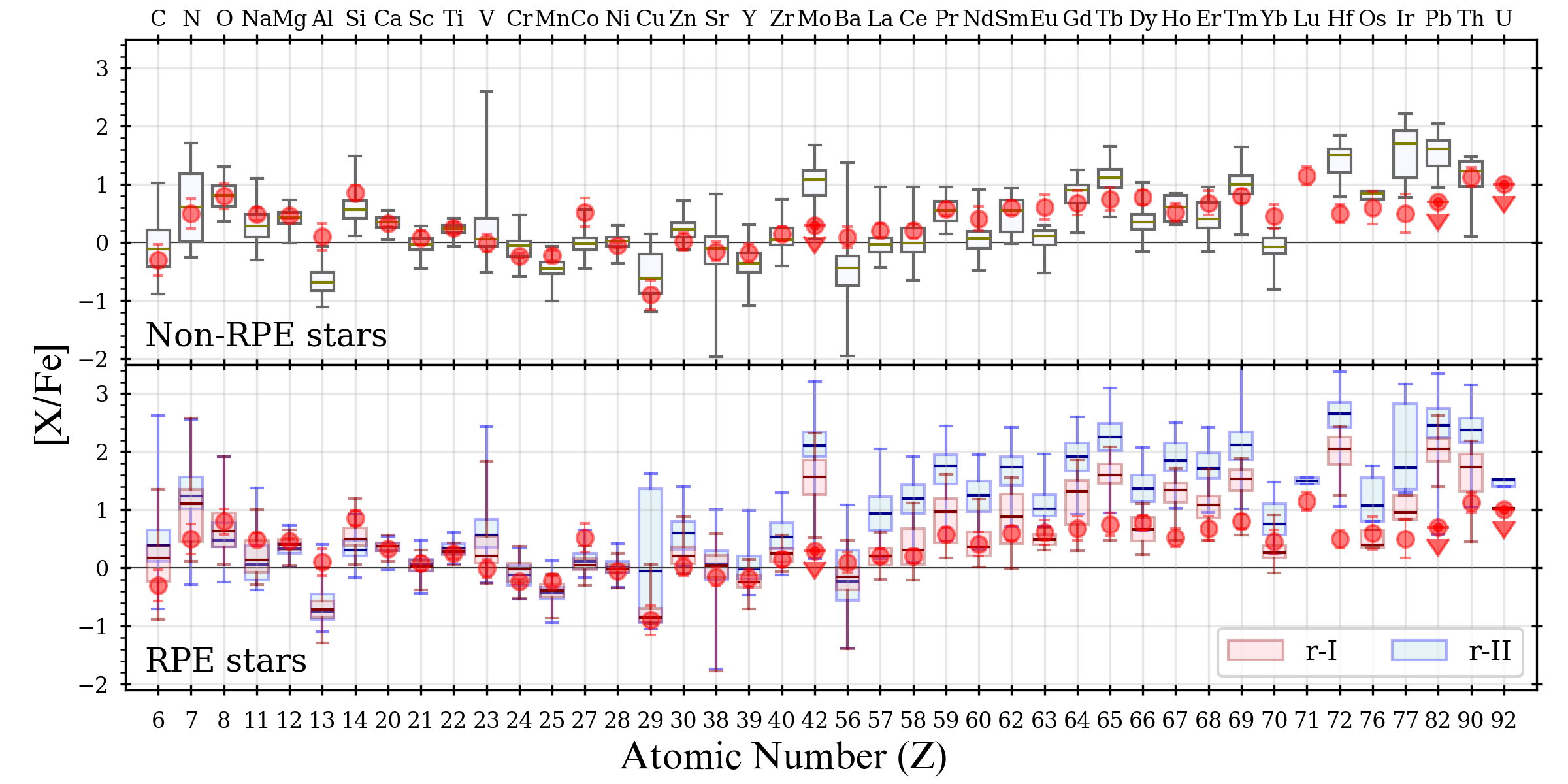} 
    \caption{Boxplot comparison of the elemental abundances ([X/Fe]) of the J1222 and 514 non-RPE and RPE stars in the metallicity range --4.0 $<$ [Fe/H] $<$ --1.0 from the SAGA database. The boxes indicate the 25th--75th percentile of the abundance distribution, and the whiskers represent the 1st--99th percentiles. Abundances and their error bars for J1222 are shown in red. Note that the abundances of Ti and V are the averages of their neutral (Ti~I, V~I) and ionized (Ti~II, V~II) species. The top panel is for the 200 non-RPE stars, while the bottom panel is for the 314 RPE stars\\. 
    }
    \label{fig_box}
\end{figure*}

To understand the distinctive abundance pattern of J1222, we compared it with stars from the SAGA database \citep{suda_stellar_2008,suda_stellar_2011}. For each star, we adopted [Fe/H]$_{\mathrm{saga}}$ as the average of [Fe/H], [Fe~I/H], and [Fe~II/H], and similarly derived [X/H]$_{\mathrm{saga}}$ for an element X by averaging [X/H], [X~I/H], and [X~II/H]. Since the database does not directly provide [X/Fe], we calculated it as [X/Fe]$_{\mathrm{saga}}$ = [X/H]$_{\mathrm{saga}}$ -- [Fe/H]$_{\mathrm{saga}}$. Note that we dropped the subscript ``saga'' in our analysis. Figure \ref{fig_box} compares [X/Fe] for J1222 with non-RPE stars (top) and RPE stars (bottom). Because the number of stars with available abundance measurements differs by element, and the abundance scatter also varies, the box sizes are not uniform, especially for neutron-capture elements. Red symbols denote our derived values. 

The top panel of Figure \ref{fig_box} indicates that, although the light-element abundance pattern of J1222 resembles that of non-RPE stars ([Eu/Fe] $\leq +0.3$), the neutron-capture elements exhibit moderate enrichment compared to that of non-RPE stars. The bottom panel compares with RPE stars, with $r$-I stars (+0.3 $<$ [Eu/Fe] $\leq$ +0.7) shown in red, and $r$-II stars ([Eu/Fe] $>$ +0.7) in blue. The abundance pattern of our program star with [Eu/Fe] = +0.61 is well correlated with the $r$-I group, showing consistent enhancements in Eu, La, Ce, Nd, and other lanthanides. Its [Ba/Eu] $\sim$ --0.5 also supports pure \rp origins, as discussed in Section \ref{sec:4.1}.

\subsection{Association of J1222 with Known Substructures}\label{sec:4.3}
Beyond its chemical composition, J1222's orbital properties offer key insights into its formation environment. To consider this, we examine J1222's dynamical characteristics and assess its potential association with known Galactic substructures.

To achieve this, we first derived the space velocity components of J1222 using 6-D astrometric parameters (position, heliocentric RV, proper motions, and distance), as listed in Table \ref{tab:tab1}. When calculating the velocity components, we adopted the Solar position $R_\odot$ = 8.2 kpc \citep{bland-hawthorn_galaxy_2016}, $Z_\odot$ = 28.8 pc \citep{bennett_vertical_2019}, Solar peculiar motion $(U, V, W)_\odot$ = (--11.0, 12.24, 7.25)\,\kms \citep{schonrich_local_2010}, and the local standard of rest velocity (LSR) $V_\text{LSR}$ = 235$\pm$3 \kms \citep{kawata_galactic_2019}. We then computed the orbital parameters of a star by performing orbital integration using the Action-based GAlaxy Modelling Architecture \texttt{AGAMA}\footnote{\url{http://github.com/GalacticDynamics-Oxford/Agama}} package \citep{vasiliev_agama_2019} with the gravitational potential model by \cite{mcmillan_mass_2017}. Note that, since we adopted a left-handed Galactic coordinate system, where $U$ is positive toward the Galactic center, we reversed the sign of $U_\odot$ from the original value provided by \citet{schonrich_local_2010}. A summary of the derived kinematic and dynamical parameters of J1222 is listed in Table \ref{tab:tab1}.

Compared with \citet{ye_dynamical_2023}'s analysis of LAMOST stars, J1222 exhibits strongly retrograde motion ($J_{\phi}$ = 2.2 $\times$ $10^{3}$~kpc~km~$s^{-1}$) and moderately high eccentricity ($e=0.72$), matching the I'itoi-Sequoia-Arjuna (ISA) group criteria. Its low metallicity ([Fe/H] = --2.45) further aligns it with the I'itoi substructure, rather than Sequoia or Arjuna. Thus, we conclude that J1222 is likely associated with the I'itoi substructure. As a member of I'itoi, the chemical-abundance pattern of J1222 can provide a valuable opportunity to study the chemical evolution of this substructure, as well as the formation environment of J1222. We discuss this in more detail in Section~\ref{sec:5.3}.

%two column
% FIGURE 6

%%%%%%%%%%%%%%%%%%%%%%%%%%%%%%%%%%%%%%%%%%%%%%%%%%%%%%%%%%%%%%%%%%%%%%%%%%%%%%%%%%%%%%%%%%%%%%%%%%%%%%%%%
%\vskip 1cm
\section{Discussion} \label{sec:5}
\subsection{Possible Sources of $R$-process Enhancement for J1222}\label{sec:5.1}
J1222 exhibits moderate \rp enhancement and a significant actinide boost. To explore the origin of this pattern, we examine whether it can be explained by specific \rp production sites, focusing on NSMs and BH-NSMs.

\vskip 1cm
\subsubsection{Neutron Star Merger}
NSMs are considered promising sites for producing heavy \rp elements. We assess whether existing NSM models, particularly those capable of reproducing actinide-rich abundance patterns, can explain the observed chemical signatures of J1222.

In NSMs, two major ejection mechanisms contribute to \rp nucleosynthesis: the dynamical ejecta \citep{kruger_estimates_2020} and the disk wind ejecta \citep{dietrich_multimessenger_2020}. The dynamical ejecta, characterized by extremely low electron fractions ($Y_{\rm e}$ = $1/(1 + N_{\rm n}/N_{\rm p}$)), where $N_{\rm n}$ is the number of neutrons and $N_{\rm p}$ is the number of protons, provides a highly neutron-rich environment favorable to actinide production. In contrast, disk wind outflows have a relatively higher $Y_{\rm e}$, contributing less to actinide synthesis. Therefore, the observed actinide enhancement is critically dependent on the relative contributions of these two components of the ejecta.

\citet{holmbeck_reconstructing_2021} performed NSM simulations under various nuclear equations of state (EOS), using observed abundance ratios such as [Fe/H], [Eu/Fe], [Zr/Dy], and [Th/Dy] to constrain the models. They demonstrated that diverse \rp abundance patterns can be reproduced by adjusting the balance between dynamical and disk wind ejecta. Among the EOSs used in the simulations, the DD2 EOS \citep{typel_composition_2010}, which describes relatively stiff nuclear matter and results in neutron stars with larger radii, was found to explain the characteristics of actinide-boost stars best. The stiffness promotes the temporary survival of hyper-massive neutron stars (HMNSs) after mergers, allowing for the prolonged ejection of neutron-rich, low-$Y_{\rm e}$ material. 

Among the stars analyzed, we identify the $r$-I actinide-boost star CS~30306-132 \citep{honda_spectroscopic_2004} as a particularly relevant comparison case. This star exhibits neutron-capture abundance ratios ([Fe/H] = --2.41, [Th/Dy] = +0.39, [Zr/Dy] = --0.40) closely similar to those of J1222. Based on this close chemical similarity, we infer that J1222 likely underwent a nucleosynthetic history analogous to that of CS~30306-132. To assess whether such a nucleosynthetic scenario is physically plausible, we briefly summarize the NSM parameters adopted in the DD2-based simulation for CS~30306-132. In the simulation of \citet{holmbeck_reconstructing_2021}, the merger involved neutron stars with masses of 1.44 and 1.40 \msun, yielding a total pre-merger mass of 2.84 \msun. The components of the mass ejected were estimated as $M_{\rm dyn} = 5.09 \times 10^{-3}$ \msun\ for the dynamical ejecta and $M_{\rm wind} = 39.88 \times 10^{-3}$ \msun\ for the disk wind ejecta, resulting in a remnant mass ($M_{\rm rem}$) of 2.79 \msun.

According to \citet{bauswein_prompt_2013}, the mass of the prompt-collapse threshold ($M_{\rm thres}$) for the DD2 EOS is 3.35 \msun. Since the total pre-merger mass (2.84 \msun) of the progenitor of CS~30306-132 is below this threshold, the remnant temporarily survives as an HMNS, allowing substantial disk wind ejection before eventual collapse. Although the progenitor had a nearly symmetric mass ratio ($M_1/M_2 = 1.03$), leading to relatively less dynamical ejecta, the low $Y_{\rm e}$ conditions maintained during the merger favored actinide production.

Consequently, the strong actinide signature observed in J1222 can be explained by a similar merger scenario: the combination of neutron-rich dynamical ejecta, enhanced disk wind contributions due to the temporary HMNS phase, and the overall remnant evolution under DD2 EOS conditions. This supports the interpretation that the progenitor of J1222 experienced a nucleosynthetic pathway and merger outcome analogous to that modeled for CS~30306-132.

\subsubsection{Black Hole-Neutron Star Merger}
Another potential site of \rp nucleosynthesis is a BH-NSM. We evaluated whether such events, under specific dynamical conditions, could reproduce the abundance pattern observed in J1222.

A BH-NSM produces two distinct types of ejecta. The dynamical ejecta, expelled immediately after the merger by strong tidal forces, are typically very neutron-rich (low $Y_{\rm e} \lesssim 0.25$), and can give rise to red kilonovae due to their lanthanide-rich composition \citep{kasen_kilonova_2015}. This environment is particularly favorable for strong production of \rp-elements, including the potential synthesis of actinides. In contrast, the post-merger ejecta, released subsequently through neutrino-driven winds from the accretion disk, possess higher electron fractions ($Y_{\rm e} \gtrsim 0.25$) and tend to produce lanthanide-poor, blue kilonovae.

\citet{wanajo_actinide-boosting_2024} conducted the first quantitative study of actinide synthesis in BH-NSMs using three-dimensional magneto-hydrodynamic (MHD) simulations. They found that fission recycling is particularly efficient in regions where $Y_{\rm e} \sim 0.04$--$0.06$ within the dynamical ejecta, leading to enhanced actinide production and elevated Th/Eu ratios. Meanwhile, the post-merger ejecta, with $Y_{\rm e} \sim 0.2$--$0.4$, contribute minimally to lanthanide and actinide synthesis.

According to their simulations, to reproduce the abundance pattern of J1222, a scenario dominated by dynamical ejecta with low-$Y_{\rm e}$ conditions is required. To support this, the Q4B5H-DD2 model presented by \citet{wanajo_actinide-boosting_2024} shows that dynamical ejecta can account for 56--69\% of the total ejecta, confirming the feasibility of such a scenario in simulations.

Although the two BH-NSMs of GW200105 and GW200115 \citep{abbott_observation_2021} were detected by gravitational-wave signals, no electromagnetic counterpart (kilonova) was observed, leaving the presence of actinide-rich ejecta unconstrained in these events. Recently, \citet{paek_gecko_2025} conducted follow-up observations of the S230518h event, modeling a range of ejecta compositions and providing valuable observational constraints. Future observations of BH-NSM kilonovae will be crucial to constrain further the formation pathways of moderately \rp-enhanced, actinide-boost stars like J1222.

\subsection{Chemical Characteristics of I'itoi Stars}\label{sec:5.2}
Although the dynamical properties of J1222 suggest a likely association with the I'itoi substructure, a larger sample is necessary to gain meaningful insights about its chemical evolution. Therefore, we searched the SAGA database for additional stars with the kinematics of the I'itoi substructure using the criteria defined by \citet{ye_dynamical_2023}, and found nine matches. However, we only used seven of these in our analysis, those having sufficiently detailed chemical-abundance measurements, including neutron-capture elements. Although the resulting sample remains small, it provides a more reliable view of the chemical characteristics of the I'itoi group of stars than relying on a single star alone.

\begin{figure*}
    \centering
    \includegraphics[scale=0.8]{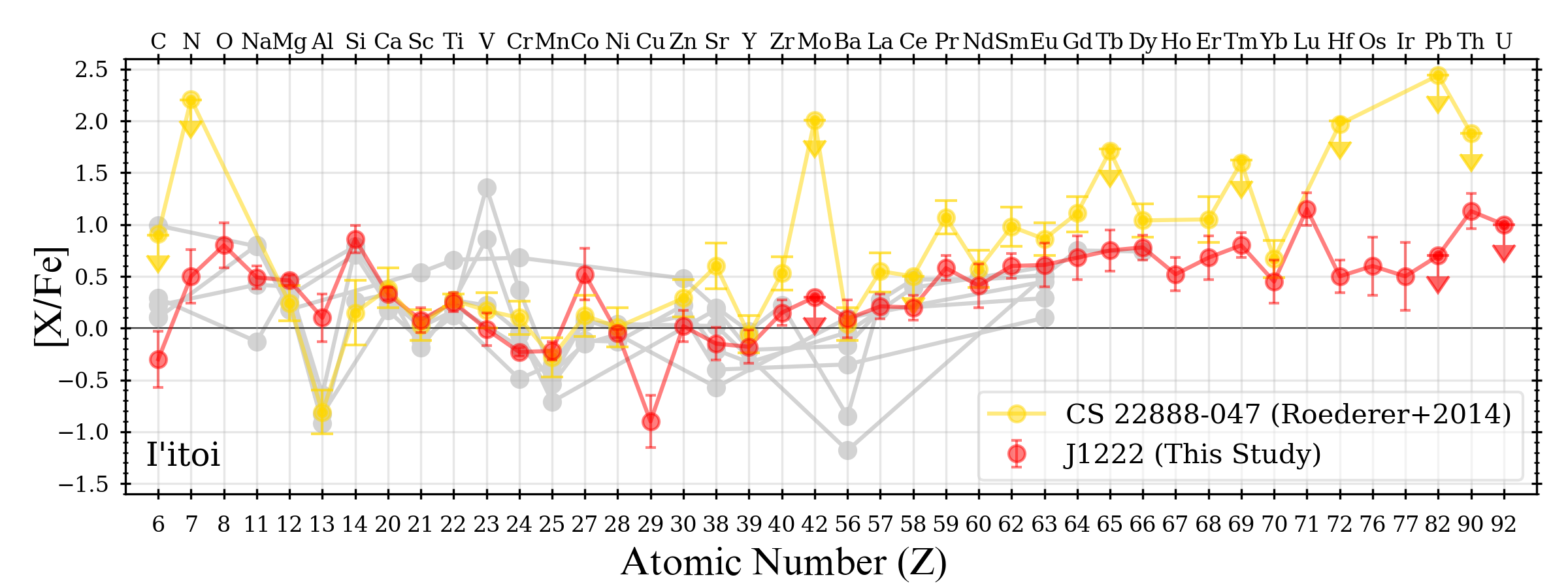}
    \caption{Chemical-abundance patterns of eight stars: J1222 (red), the actinide-boost candidate CS~22888--047 (yellow), and six other stars (gray) with similar kinematics to the I'itoi substructure selected from the SAGA database. See Sections \ref{sec:4.3} and \ref{sec:5.2} for the selection of these stars.
    }
    \label{fig_iitoi_chem}
\end{figure*}

Here, we discuss the chemical-abundance patterns of eight stars (seven from SAGA and J1222), highlighting the diversity of \rp enrichment within the group. Since stars with a common origin often share similar star-formation and enrichment histories, one might expect them to exhibit relatively consistent chemical patterns{, although deviations are commonly observed in low-mass systems.} Figure \ref{fig_iitoi_chem} shows the chemical-abundance patterns of the eight I'itoi stars from SAGA and J1222. The mean metallicity of these stars, $\langle[\mathrm{Fe}/\mathrm{H}]\rangle$ = --2.51$\pm$0.31, which is consistent with the distribution in \citet{naidu_reconstructing_2021}.

%two column
% FIGURE 7

Despite their similar metallicities, as shown in Figure \ref{fig_iitoi_chem}, the stars display significant abundance dispersions in light and heavy elements. Most exhibit typical enhancements of the $\alpha$ elements, with mean values of $\langle[\mathrm{Mg}/\mathrm{Fe}]\rangle$ = +0.34$\pm$0.10, $\langle[\mathrm{Ca}/\mathrm{Fe}]\rangle$ = +0.30$\pm$0.07 and $\langle [\mathrm{Si}/\mathrm{Fe}] \rangle$ = +0.59$\pm$0.28. In contrast, the neutron-capture elements show a much larger element-to-element dispersion, with $\langle[\mathrm{Sr}/\mathrm{Fe}] \rangle$ = --0.07$\pm$0.36, $\langle [\mathrm{Y}/\mathrm{Fe}] \rangle$ = --0.16$\pm$0.11, $\langle [\mathrm{Ba}/\mathrm{Fe}] \rangle$ = --0.29$\pm$0.45, and $\langle [\mathrm{Eu}/\mathrm{Fe}] \rangle$ = +0.49$\pm$0.21. The value of [Sr/Fe] spans more than 1\,dex, ranging from Sr-poor ([Sr/Fe] $<$ --0.5) to Sr-rich ([Sr/Fe] $>$ +0.5), indicating substantial chemical inhomogeneity across the I'itoi stars.

As these stars satisfy the same kinematic criteria from \citet{ye_dynamical_2023}, \citet{horta_chemical_2023}, and \citet{naidu_evidence_2020}, we consider them to have originated from a common progenitor system. In this context, the wide abundance dispersion in neutron-capture elements reflects inhomogeneous mixing or localized enrichment, which is characteristic of typical low-mass systems \citep{frebel_near-field_2015,hirai_enrichment_2015, hirai_infall_2022, hirai_r-process_2024, jeon_connecting_2017,alexander_inhomogeneous_2023}.

All eight I'itoi stars, including J1222, have measured Eu abundances. Among them, two are non-RPE, five are $r$-I, and one is classified as an $r$-II star. Three stars exhibit signatures of light \rp enhancement, with [Sr/Ba] $>$ +0.5 and [Ba/Eu] $<$ 0, suggesting that the light and main \rp elements may have been contributed from distinct astrophysical origins \citep[e.g.,][]{travaglio_galactic_2004,hirai_enrichment_2019}.

Interestingly, we identified one actinide-boost candidate among the I'itoi stars (CS~22888--047) from a previous study \citep{roederer_search_2014}, and its kinematic association with the I'itoi substructure is newly recognized in this work. The presence of J1222 and one actinide-boost candidate within the same kinematic group reflects the diversity of \rp enrichment among these stars. In addition, the coexistence of non-RPE, $r$-I, $r$-II, and actinide-boost star among these kinematically linked stars indicates substantial variation in \rp production within the I'itoi progenitor system. This diversity underscores the complexity of the enrichment history and suggests that \rp contribution to this system was not chemically uniform. We further discuss these interpretations of this diversity in the following section.

%\vskip 1cm
\subsection{Sources of $R$-process Production in the I'itoi Substructure} \label{sec:5.3}

In this section, we evaluate the physical feasibility of \rp production events that could account for the diverse \rp abundance patterns observed among the I'itoi stars. The primary candidate scenarios considered include collapsars, magneto-rotational supernovae (MR-SNe), NSMs, and BH-NSMs. Most of these events release energetic feedback on the order of $10^{51}$--$10^{52}~erg$, which can significantly impact the structural integrity of dwarf galaxies with shallow gravitational potential wells.

\citet{naidu_live_2022} estimated the stellar mass of the I'itoi substructure to be log~($M_{\star}$/\msun) = 6.3 ($\sim$2 $\times$ $10^{6}$ \msun), based on relative star count ratios with respect to Gaia-Sausage-Enceladus (GSE; $5 \times 10^8$ \msun). By applying the stellar mass-halo mass relation from \citet{moster_constraints_2010, moster_galactic_2013}, we estimate the corresponding dark matter halo mass to be approximately $2 \times 10^9$ \msun. Further, using the scaling relation from \citet{walker_universal_2009}, we infer the half-light radius to be $\sim$100 pc and an escape velocity of $\sim$415 \kms for the system.

The typical natal kick velocity of neutron stars is $\sim$400 \kms \citep{hobbs_statistical_2005}, with some cases exceeding 1000 \kms. However, the natal kick associated with GW170817 was inferred to be only 200--250 \kms \citep{abbott_gw170817_2017, blanchard_electromagnetic_2017}. These estimates imply that the I'itoi progenitor system could retain the ejecta of energetic \rp events, particularly those involving relatively modest natal kicks or limited feedback, such as NSMs or BH-NSMs. 

As described in Section \ref{sec:5.1}, both NSMs and BH-NSMs can plausibly reproduce the \rp abundance pattern observed in J1222, provided that the ejecta are sufficiently neutron-rich and the natal kicks are moderate. Our current analysis further demonstrates that the I'itoi progenitor had sufficient mass to gravitationally retain the \rp ejecta, bolstering the feasibility of these events not only for J1222 but for the I'itoi stellar group as a whole. These results support the feasibility of \rp enrichment through NSMs and BH-NSMs within the I'itoi progenitor system.

Moreover, systems with halo masses on the order of $10^9$ \msun\ can theoretically withstand the energetic feedback associated with collapsars, MR-SNe, and BH-NSMs. However, the long-term effectiveness of gas retention, subsequent star formation, and the propagation of \rp material critically depends on the frequency, location, and mixing efficiency of enrichment events. Given this, we examine whether the remaining two scenarios (collapsars and MR-SNe) can also account for the observed chemical signatures in the I'itoi group.

\subsubsection{Collapsars}\label{sec:5.4.1}
\citet{macfadyen_collapsars_1999} introduced the collapsar model, proposing that a massive star ($>$ 30 \msun) may collapse into a black hole following the failure of a typical core-collapse supernova (CCSN). In this model, a rapidly rotating and highly magnetized core forms an accretion disk and launches bipolar jets, leading to a highly anisotropic explosion. Building upon this framework, recent magneto-hydrodynamic (MHD) simulations \citep{siegel_collapsars_2019} have demonstrated that the accretion disk in a collapsar can generate significant amounts of both light \rp elements and main \rp elements (e.g., Eu). These simulations suggest that a single collapsar could yield approximately 10--30 times more \rp material compared to a typical NSM.

However, collapsars are extremely rare events, and their occurrence rates are expected to be low, especially in typical dwarf galaxies or their remnants. Furthermore, the production of actinide elements (e.g., Th, U) in collapsar ejecta remains poorly constrained by current theoretical and numerical models. As highlighted by \citet{brauer_collapsar_2021}, earlier modeling efforts \citep{surman_nucleosynthesis_2006, miller_full_2020} suggest that collapsar outflows are typically too neutron-poor to synthesize actinides or third-peak \rp elements. Furthermore, recent resistive-magneto-hydrodynamics (RMHD) simulations presented in \citet{shibata_self-consistent_2025} support this, by showing that strong magnetic fields drive outflows before sufficient neutron capture takes place, limiting actinide production. Therefore, while the collapsar scenario provides a compelling explanation for the enrichment in light \rp elements and the observed dispersion in Eu abundances, it appears not to be able to account for the existence of actinide-boost stars.

\subsubsection{Magneto-rotational Supernova/Hypernova (MR-SN/HN)}\label{sec:5.4.2}
The MR-SN model was introduced as a variant of CCSNe involving a rapidly rotating, strongly magnetized progenitor. In this scenario, a collimated jet, launched by the magnetic fields, drives an anisotropic explosion and provides a promising site for \rp nucleosynthesis \citep{winteler_magnetorotationally_2012, nishimura_r_2015, nishimura_intermediate_2017}. This model is particularly efficient at synthesizing light \rp elements such as Sr, Y, and Zr \citep{halevi_r-process_2018}. Due to the short lifetimes of their massive progenitors, MR-SNe are expected to exhibit short delay times (on the order of a few Myr), making them plausible contributors to \rp enrichment in low-metallicity ([Fe/H] $\leq$ --2) environments \citep{siegel_collapsars_2019, cote_neutron_2019}.

\citet{winteler_magnetorotationally_2012} further proposed that prompt jet ejecta from MR-SNe with extreme magnetic fields and rapid rotation could produce heavy \rp elements, including actinides. Building on this, \citet{jha_observational_2019} interpreted the abundance patterns of some highly RPE stars as signatures of MR-SN events.

However, 3D MHD simulations by \citet{zha_nucleosynthesis_2024}, focusing on disk wind ejecta, rather than prompt jets, predicted predominantly light \rp element production ($A$ $\leq$ 130), but insignificant actinide synthesis. Similarly, \citet{nishimura_intermediate_2017} demonstrated that non-jet ejecta generally produce elements only up to the second \rp peak, with negligible actinide yields. Observationally, \citet{yong_r-process_2021} suggested that SMSS J200322.54-114203.3 may be a remnant of an MR-SN, based on the combination of [Fe/H] = --3.5, [Zn/Fe] = +0.72, enhanced light \rp elements, and a notable lack of actinides.

The I'itoi stars analyzed in this study do not consistently exhibit enhancements in light \rp elements (e.g., Sr or Zn), with only a subset of stars displaying elevated [Sr/Fe]. Furthermore, the diverse phenomenology observed -- including $r$-I, $r$-II, actinide-boost, and both Sr-rich and Sr-poor stars -- is difficult to reconcile with a single MR-SN origin. Consequently, while MR-SNe are unlikely to be the dominant source of \rp material in this system, they may have contributed to the enrichment of some Sr-rich stars, or acted as secondary sources of light \rp nucleosynthesis.

\subsection{Temporal and Dynamical Constraints on $R$-process Enrichment} \label{sec:5.2a} 
The U/Th chronometric age of J1222 provides a conservative lower bound on the timescale over which the \rp-enriched gas environment from the star formed must have been maintained long enough to enable star formation. In this sense, the chronometer constrains the duration of the chemically enriched gas supply rather than directly tracing the timing of the \rp event itself. 

Because the uranium abundance in this work is constrained only by an upper limit, the derived U/Th age does not represent a precise absolute age. Instead, it reflects the minimum time that has elapsed since the actinides were last effectively synthesized, during which the gas should have remained relatively unmixed and undiluted. This is consistent with previous studies demonstrating that, in actinide-boost stars, the U/Th ratio is more reliable as a constraint on the duration of chemically unusual star-forming environments than as a precise chronometer of individual \rp events \citep{cayrel_uranium_2001,holmbeck_r-process_2018,roederer_r-process_2022,roederer_fission_2023}.

This is particularly relevant for low-mass, chemically inhomogeneous systems such as I'itoi progenitor, which were later accreted into and disrupted within the MW. During infall, processes such as tidal stripping, dynamical heating, orbital evolution, and ram-pressure stripping accelerate gas dilution and mixing, making it difficult to preserve locally formed extreme \rp signatures over extended timescales \citep{hirai_infall_2022}. Numerical simulations indicate that these processes rapidly erode chemically coherent star-forming environments in dwarf galaxies after accretion \citep{mayer_rampressure_2006,nichols_rampressure_2011,fillingham_stripping_2016,hirai_r-process_2024}. In this context, the key question shifts from how many \rp events occurred to how long the products of such events could be retained in environments stable enough to form stars like J1222.

The presence of Sr-rich stars in the I'itoi system does not require multiple actinide-producing events. Light \rp elements such as Sr, Y, and Zr can be synthesized in a variety of astrophysical environments, without significant actinide production, and they show much larger star-to-star dispersions than heavy \rp elements \citep[e.g.,][]{honda_HD122563_2006,roederer_ubiquity_2010,hansen_rprocess_2018}. This behavior suggests that light and heavy \rp nucleosynthesis can proceed through partially decoupled pathways \citep{qian_lightr_2008,hansen_AgPd_2012}, indicating that the presence of Sr-rich stars alone does not tightly constrain the nature or number of actinide-producing enrichment events.

Taken together, the U/Th chronometric age constrains the minimum timescale over which the I'itoi progenitor system must have maintained an environment capable of producing chemically extreme stars before its collapse. This places J1222 within a broader framework in which \rp enrichment, star formation, and substructure assembly are understood as interconnected aspects of the early assembly of the MW. This time constraint is consistent with a scenario in which J1222 formed after \rp enrichment but before the I'itoi progenitor was accreted into the Milky Way and underwent significant chemical and dynamical disruption.

%\vskip 2cm
%% CONCLUSION: Section 6 %%
\section{Conclusions and Future Work}\label{sec:6}
We have analyzed the detailed chemical-abundance pattern of LAMOST J122216.85-063345.2 (J1222) using high-resolution spectra acquired during the system verification (SV) run of Gemini-S/GHOST. A total of 47 elemental abundances were determined. The light elements exhibit features consistent with typical halo stars. However, in heavy elements, especially those associated with neutron capture, J1222 exhibits moderate \rp enhancement and actinide-boost characteristics. 

We examine possible enrichment scenarios involving NSMs and BH-NSMs to interpret this unusual heavy-element pattern. The dynamical properties of J1222 are well-aligned with the I'itoi substructure defined by \citet{naidu_evidence_2020}, \citet{horta_chemical_2023},  and \citet{ye_dynamical_2023}. We performed a detailed chemical analysis of seven stars with similar kinematics to the I'itoi structure selected from the SAGA database, together with J1222. We found that most I'itoi stars exhibited alpha-element enhancement, consistently low [Ba/Fe], and a large dispersion in [Eu/Fe]. These features suggest that the \rp enrichment within the progenitor system may not have originated from a single event, but rather varied spatially and/or temporally.

In addition, we explored several astrophysical scenarios to explain the observed \rp abundances. Our analysis suggests that NSMs and BH-NSMs were likely the main contributors to the enrichment, while MR-SNe may have played a secondary role in enriching some light \rp element (e.g., Sr)-rich stars. 

Notably, J1222 and other actinide-boost candidate (CS 22888-047) are identified among stars kinematically associated with the I'itoi substructure. The coexistence of non-RPE, $r$-I, $r$-II, and an actinide-boost star within the same kinematic association highlights the diversity of \rp enrichment in the I'itoi progenitor system. Moreover, this coexistence, together with the wide dispersion observed in light \rp element abundances (e.g., Sr), suggests that \rp enrichment in this system was not chemically uniform and likely involved localized enrichment and inefficient mixing.

In addition to identifying the astrophysical origin of \rp enrichment, the U/Th chronometer constrains the timescale during which the progenitor system retained enriched gas capable of forming stars such as J1222. The U/Th upper limit indicates an age range of approximately 9 to 11 Gyr, suggesting that J1222 formed after the \rp event but before the final accretion and disruption of its progenitor system. This constraint does not provide a precise stellar age. Instead, it outlines a temporal window during which enriched material was preserved and contributed to subsequent star formation. As such, it offers a chemically informed timescale that helps connect \rp enrichment, star formation, and assembly of halo substructures in the early MW.

Several I'itoi-associated stars listed in the SAGA database remain unexplored with high-quality, high-resolution spectroscopy, particularly with respect to their heavy-element abundance patterns. Detailed chemical characterization of these stars will be essential for reconstructing the chemical-enrichment history and internal diversity of the I'itoi progenitor system. In combination with temporal constraint from J1222, such system-wide abundance patterns offer complementary perspective on how $r$-process enrichment, subsequent star formation, and inefficient mixing proceeded before the system's accretion into the MW. Ultimately, we anticipate that systematic high-resolution chemical studies of kinematically associated stars will place the I'itoi system in a broader context of early Galactic assembly and chemical evolution.

%Facility
\facilities{Gemini:South (GHOST)}

%Software
\software{
{\texttt{DRAGONS}}\,\citep{labrie_dragonsquick_2023},
{\texttt{IRAF}}\,\citep{tody_iraf_1986,tody_iraf_1993,fitzpatrick_iraf_2025}, 
{\texttt{LINEMAKE}}\,\citep{placco_linemake_2021,placco_linemake_2021-1},
{\texttt{MOOG}}\,\citep{sneden_carbon_1973},  
{\texttt{Numpy}}\,\citep{harris_array_2020}, 
{\texttt{Pandas}}\,\citep{mckinney_data_2010}, 
{\texttt{Astropy}}\,\citep{astropy_collaboration_astropy_2022}, 
}
\clearpage
\section*{Acknowledgments}
%\onecolumngrid
%\begin{acknowledgments}

The authors are grateful to the anonymous referee for the constructive comments and suggestions that significantly improved the clarity and quality of this manuscript. The authors also thank Dr. Ian Roederer for his valuable comments and insightful suggestions. The authors thank Christian R.\ Hayes, Kim Venn, Siyi Xu, David Henderson, Pablo Prado, Carlos Quiroz, Chris Simpson, Cristian Urrutia, Trystyn Berg, Michael Ireland, Alan McConnachie, John Pazder, Fletcher Waller, John Blakeslee, Zachary Hartman, Bryan Miller Cao, Janice Lee, David O. Jones, and Susan Ridgway for their contributions to the GHOST System Verification (SV).

Y.S.L. acknowledges support from the National Research Foundation (NRF) of Korea grant funded by the Ministry of Science and ICT (RS-2024-00333766). The work of V.M.P. is supported by NSF NOIRLab, which is managed by the Association of Universities for Research in Astronomy (AURA) under a cooperative agreement with the U.S. National Science Foundation. Y.H.\ acknowledges support from the JSPS KAKENHI Grant Numbers JP22KJ0157, JP25H00664, and JP25K01046. T.C.B. acknowledges partial support from grants PHY 14-30152; Physics Frontier Center/JINA Center for the Evolution of the Elements (JINA-CEE), and OISE-1927130; The International Research Network for Nuclear Astrophysics (IReNA), awarded by the US National Science Foundation, and DE-SC0023128CeNAM; the Center for Nuclear Astrophysics Across Messengers (CeNAM), awarded by the U.S. Department of Energy, Office of Science, Office of Nuclear Physics. E.M. acknowledges funding from FAPEMIG under project number APQ-02493-22 and a research productivity grant number 309829/2022-4 awarded by the CNPq, Brazil. This work was also supported by the Korea Astronomy and Space Science Institute (KASI) grant funded by the Korean government (No. 2026-186-000, International Optical Observatory Project). 

Based on observations obtained at the international Gemini Observatory, a program of NSF NOIRLab GS-2023A-SV-101, which is managed by the Association of Universities for Research in Astronomy (AURA) under a cooperative agreement with the U.S. National Science Foundation on behalf of the Gemini Observatory partnership: the U.S. National Science Foundation (United States), National Research Council (Canada), Agencia Nacional de Investigaci\'{o}n y Desarrollo (Chile), Ministerio de Ciencia, Tecnolog\'{i}a e Innovaci\'{o}n (Argentina), Minist\'{e}rio da Ci\^{e}ncia, Tecnologia, Inova\c{c}\~{o}es e Comunica\c{c}\~{o}es (Brazil), and Korea Astronomy and Space Science Institute (Republic of Korea). Data processed using DRAGONS (Data Reduction for Astronomy from Gemini Observatory North and South). GHOST was built by a collaboration between Australian Astronomical Optics at Macquarie University, National Research Council Herzberg of Canada, and the Australian National University, and funded by the International Gemini partnership. The instrument scientist is Dr. Alan McConnachie at NRC, and the instrument team is also led by Dr. Gordon Robertson (at AAO), and Prof. Michael Ireland (at ANU). The authors would like to acknowledge the contributions of the GHOST instrument build team, the Gemini GHOST instrument team, the full SV team, and the rest of the Gemini operations team that were involved in making the SV observations a success.
%\end{acknowledgments}
%\twocolumngrid

%% Reference %%
%\bibliographystyle{aasjournal}
\bibliographystyle{aasjournalv7}
%\begin{bibliography}{refer_mjeong_may23.bib}
%\end{bibliography}

%\bibliographystyle{aasjournal}
%\bibliographystyle{apj}
\bibliography{test_ref}{}

\appendix
%\section{Line List used in the Abundance Analysis}
\renewcommand{\thetable}{A\arabic{table}}
\setcounter{table}{0}
Table \ref{tab:linelist} summarizes the spectral lines used in the abundance analysis, listing the adopted atomic parameters (wavelength, excitation potential, and oscillator strength), the derived abundances for individual lines, upper-limit indicators where applicable, and the analysis method (equivalent width based analysis or a spectral synthesis based analysis).
\begin{deluxetable*}{cccccccc}[htp!]
\tablecaption{Line List used in the Abundance Analysis\label{tab:linelist}}
\tablewidth{0pt}
\tabletypesize{\scriptsize}
\tablehead{
\colhead{Species} &
\colhead{Ion} &
\colhead{Wavelength (\AA)} &
\colhead{E.P (eV)} &
\colhead{log $gf$} &
\colhead{$\log\epsilon$(X)} &
\colhead{Limit flag} &
\colhead{Method$^{a}$}
}
\startdata
Li   & I &  6707.80 & 0.00 &  ~~0.17  & --0.60 & $<$ & S\\
CH   & \nodata &  4313.00 & \nodata  & \nodata   &  ~~5.53 & 0 & S\\
CN   & \nodata &  3875.00 & \nodata  & \nodata   &  ~~5.88 & 0 & S\\
$[$\ion{O}{1}$]$ & - &  6300.80 & 0.00 & --9.69  &  ~~0.80 & $<$ & S\\
Na   & I &  5889.95 & 0.00 &  ~~0.11  &  ~~4.24 & 0 & E\\
Na   & I &  5895.92 & 0.00 & --0.19  &  ~~4.34 & 0 & E\\
\enddata
\tablecomments{
$^{a}$The method used to derive the abundance is indicated as follows: `E' denotes equivalent width analysis, while `S' indicates spectral synthesis. This table is available in its entirety in machine-readable format. In the `Limit flag' column, `<' indicates an upper limit, while `0' denotes a detected abundance.}
\end{deluxetable*}

\end{CJK}
\end{document}